\documentclass[11pt,a4paper]{article}
\usepackage{jheppub}
\usepackage{mathrsfs,graphicx,rotating,amsmath,amsfonts,mathtools,booktabs,amssymb,wasysym,caption}
\usepackage{hyperref}
\usepackage{slashed}
\usepackage[table,xcdraw,dvipsnames]{xcolor}
\usepackage{graphicx}
\usepackage{bbold}
\usepackage[utf8x]{inputenc}
\usepackage[english]{babel}
\usepackage{multirow,multicol}
\usepackage{epstopdf}
\usepackage{bbm}
\usepackage{changepage}
\usepackage{appendix}
\usepackage{libertine}
\usepackage{braket}
\usepackage{wasysym}
\usepackage{empheq}
\usepackage{cancel}
\usepackage{enumitem}
\usepackage{mathrsfs}
\usepackage{lmodern}
\usepackage{tabularx}
\usepackage{multicol}
\usepackage{color}
\usepackage{mathtools}
\usepackage{verbatim}

\newcommand{\mc}{\mathcal}
\newcommand{\be}{\begin{equation}}
\newcommand{\ee}{\end{equation}}
\newcommand{\bea}{\begin{eqnarray}}
\newcommand{\eea}{\end{eqnarray}}
\newcommand{\half}{\frac{1}{2}}

\setcitestyle{square}

\DeclarePairedDelimiter\abs{\lvert}{\rvert}%
\title{Moduli Stabilisation and the Holographic Swampland}
\author[a]{Joseph P. Conlon,} \author[a]{Filippo Revello}
\affiliation[a]{Rudolf Peierls Centre for Theoretical Physics\\ Beecroft Building, Clarendon Laboratory, Parks Road, University of Oxford, OX1 3PU, UK}
\emailAdd{joseph.conlon@physics.ox.ac.uk}\emailAdd{filippo.revello@physics.ox.ac.uk}
\abstract{We investigate whether Swampland constraints on the low-energy dynamics of weakly coupled, moduli-stabilised string vacua in AdS can be related to inconsistencies of their putative holographic duals or, more generally, recast in terms of CFT data. We find that various swampland consistency constraints are equivalent to a negativity condition on the sign of certain mixed anomalous dimensions. This condition is similar to well-established CFT positivity bounds arising from causality and unitarity, but not known to hold in general. The studied scenarios include LVS, KKLT, and both perturbative and racetrack stabilisation. Interestingly, the LVS vacuum (with $\Delta_{\varphi} = 8.038$) also appears to live very close to a critical value ($\Delta_{\varphi} = 8$) where the anomalous dimensions change sign. We finally point out an intriguing connection to the Swampland Distance Conjecture, both in its original and refined versions.}
\begin{document}
\maketitle
\section{Introduction}\label{sc:intro}
The last few years have seen a resurgence of interest in the identification of criteria that distinguish between the low-energy effective theories which admit an ultraviolet completion in String Theory, or more generally within the framework of Quantum Gravity, from those which do not. The former are commonly referred to as the $
\emph{Swampland}$ \cite{Vafa:2005ui,Ooguri:2006in}, as opposed to the $\emph{Landscape}$ \cite{Susskind:2003kw} of acceptable theories (see \cite{Palti:2019pca} for a comprehensive review). The underlying idea for the emergence of such constraints is that self-consistency conditions should become more and more restrictive as one moves towards the UV, culminating with string theory whose structure at high energies is essentially determined.

This program has resulted in a large number of conjectures which vary greatly in their predictive power and their degree of
rigourous support. These are often linked by a deep web of connections. Some of them, such as the No Global Symmetries \cite{Banks:1988yz, Banks:2010zn} and Weak Gravity conjectures \cite{ArkaniHamed:2006dz}, have existed for some time and have a high level of support, relying both on a UV stringy logic and also IR arguments involving black hole physics. On the other end of the spectrum, there are speculative hypotheses such as the dS Conjecture \cite{Obied:2018sgi}, which are motivated by the difficulties encountered in the construction of explicit counterexamples in String Theory and have less rigorous support, but carry broad phenomenological implications \cite{Obied:2018sgi,Agrawal:2018own}.

With a somewhat analogous philosophy, but using very different methods, the Bootstrap program has been working in the
direction of determining restrictions on consistent Conformal Field Theories (CFTs) based only on the assumption of certain fundamental requirements such as unitarity and crossing symmetry (an introduction to the topic can be found in \cite{Simmons-Duffin:2016gjk}).
This has had considerable success in carving out the parameter space of consistent CFTs through a clever combination of numerical \cite{Rattazzi:2008pe,ElShowk:2012ht} and analytical techniques \cite{Komargodski:2012ek,Fitzpatrick:2012yx}, showing that certain interesting theories (such as the Ising model) live near the boundary of the allowed regions.

The underlying theme of this paper is to explore whether or not these two programs, so different in principle, can be connected through the AdS/CFT correspondence.
The attractiveness of such an approach is twofold. First, AdS/CFT (at least in its strongest formulation) can be thought to apply to generic theories of quantum gravity and therefore be more general than arguments based on weakly coupled string theory. Second, it may provide an alternative, possibly more fundamental, perspective on the swampland conjectures. The low-energy effective Lagrangians for string compactifications often arise after a series of several steps of dimensional reduction.  By necessity, these steps leading to $\mc{N}=0$ minima of the effective potential
rely on perturbative expansions and weak coupling approximations. AdS/CFT may provide an alternative means to test the correctness of these ideas.

A number of recent works have tried to exploit this rather broad idea (see also the discussion in section 3.3.5 of the review \cite{Palti:2019pca}).
Examples include general arguments in support of the Weak Gravity Conjecture \cite{150901647, Benjamin:2016fhe,Montero:2016tif,170909159,Urbano:2018kax,Montero:2018fns} and for the absence of global symmetries in quantum gravity theories with a holographic dual \cite{Harlow:2018tng,Harlow:2018jwu}.

Here our approach aims at analysing the properties of the specific CFTs which would be dual to weakly coupled, effective theories arising from string compactification scenarios. Our focus is therefore on models with $\mathcal{N} \le 1$ supersymmetry and potentials for moduli. Although the exploration of CFT duals to weakly-coupled low energy theories of moduli stabilisation has been the subject of occasional studies \cite{Aharony:2008wz,deAlwis:2014wia, Conlon:2018vov}, the topic remains relatively unexplored, in part due to the belief that dual CFTs would have extremely complex properties.  

However, it was found in \cite{Conlon:2018vov} that
the Large Volume Scenario \cite{Balasubramanian:2005zx,Conlon:2005ki}, viewed holographically, admits a relatively simple and unique form for the low-lying operators and their conformal dimensions. This makes it a relatively well-posed question whether Swampland-like constraints on the
corresponding low energy Lagrangians can be recast in the language of conformal theories.

The aim of this paper is to make these statements quantitative.
In this paper we focus only on AdS constructions and do not discuss subsequent uplifts to de Sitter space -- although there are also
interesting conceptual and consistency questions associated to de Sitter constructions, the use of AdS/CFT techniques restricts our analysis to the first step involving AdS vacua, which in any case already have many phenomenologically interesting features.

In general, one complication is that these Lagrangians can contain a large number of light moduli $\varphi_i$, typically of order a few hundred (arising from typical Calabi-Yau values for $h^{1,1}$ and $h^{2,1}$). In a holographic picture, each particle in the bulk corresponds to a single trace primary operator in the CFT$_d$ of conformal dimension $\Delta$ given by
\begin{equation}\label{eq:dim}
\Delta (\Delta-d) = m^2 R^2_{AdS}.
\end{equation}
For `generic' $\mc{N}=1$ supergravity models, $V_{AdS} \sim -m_{3/2}^2 M_P^2, R_{AdS} \sim m_{3/2}^{-1}$ and $m_{\varphi_i} \sim m_{3/2}$,
translating into a large number of operators of low conformal dimension.

For this reason we shall mostly focus on a specific case, the Large Volume Scenario (LVS), which allows for a much simpler holographic description \cite{Conlon:2018vov}. For LVS in the large volume limit, the only operators to retain a small $\mc{O}(1)$ conformal dimension $\Delta$  are the massless graviton, the volume modulus and its axionic partner; all other dimensions diverge and the corresponding fields can be integrated out. Above these, there is a large $\Delta_{gap}$, not just to the higher spin states but also to the other moduli. Furthermore, the limit of infinite $\mathcal{V}$ also provides a well defined and universal limit, where all the interactions in the effective theory are fixed and do not depend on the fine details of the compactification.

This paper is organized as follows: after the introduction, we first review the basic aspects of the LVS construction in section \ref{sc:lvs} and the properties of the putative holographic dual. As our analysis focuses on potential changes of signs in the low-energy AdS Lagrangian we then examine positivity bounds, both within CFTs and from the S-matrix, along with a brief review of some analytic bootstrap techniques in section \ref{sc:boot}. In section \ref{sc:app}, we propose a Swampland criterion based on the large-$\ell$ sign of mixed anomalous dimensions in the CFT and show that this is capable of reproducing various swampland results including those on axion field ranges and also (in section \ref{sec5}) on the refined swampland distance conjecture. We also discuss the application of this criterion to other related scenarios of moduli stabilisation such as KKLT, racetrack and perturbative stabilisation. Finally, possible future directions and a few unresolved questions are presented in section \ref{sc:conc}.

\section{The Large Volume Scenario and its Holographic Properties}\label{sc:lvs}

We start with a brief review of the Large Volume Scenario of moduli stabilisation (LVS), focusing on properties of both the AdS vacuum and its putative holographic dual CFT.

\subsection{Introduction to LVS and its low energy dynamics}
The Large Volume Scenario \cite{Balasubramanian:2005zx,Conlon:2005ki} is a particular model of IIB compactification with fluxes, where all moduli are stabilized in a susy-breaking AdS vacuum at an exponentially large value of the volume $\mathcal{V}$. This is achieved through a combination of perturbative corrections arising from the $\alpha'^3\mathcal{R}^4$ term and non-perturbative effects in the superpotential.
A pedagogical account of the basic construction is given in \cite{hepth0611039}. The dilaton and complex structure moduli are heavy and stabilised by fluxes. These can then be integrated out, leading to an effective theory described by a Kähler potential $K$ and superpotential $W$ of the form
\begin{equation}\label{eq:k}
K = -2 \log \Bigg( \mathcal{V} + \frac{\xi}{g_s^{3/2}} \Bigg),
\end{equation}
\begin{equation}
W = W_0+\sum_{i} A_i e^{-\alpha_i T_i},
\end{equation}
which depends on the Kähler moduli
$T_i = \tau_i + i a_i$.
The coefficient $\xi$ of the $\alpha'^3$ correction is determined by the Euler characteristic of the compactification manifold,
$\xi = \frac{\zeta(3) \chi(M)}{2 (2 \pi)^3}$, while the $\alpha_i$ in the exponential can assume the values $ \alpha_i= \frac{2 \pi}{N}$.

In the simplest scenario (which we restrict to here), there are only two Kähler moduli,
one large cycle corresponding to the size of the overall volume ($T_b$) and one small internal `blow-up' cycle ($T_s$). The volume is then written as
\begin{equation}
\mathcal{V}= \frac{1}{\kappa} \big(\tau_b^{3/2}- \tau_s^{3/2}\big),
\end{equation}
with the constant $\kappa$ depends on the specific Calabi-Yau.
Inserting this expression into (\ref{eq:k}), one can use the form of the $\mathcal{N}=1$ supergravity potential
\begin{equation}
V= e^{K} \big(G^{T \bar{T}} D_{T} W \bar{D_{T}} W - 3 \abs{W}^2 \big),
\end{equation} to obtain an effective potential
\begin{equation}
V=\frac{A a_{s}^{2} \sqrt{\tau_{s}} e^{-2 a_{s} \tau_{s}}}{\mathcal{V}}-\frac{B W_{0} a_{s} \tau_{s} e^{-a_{s} \tau_{s}}}{\mathcal{V}^{2}}+\frac{C \xi W_{0}^{2}}{g_{s}^{3 / 2} \mathcal{V}^{3}},
\end{equation}
where $A$, $B$ and $C$ are numerical constants,
which is minimized for
\begin{equation}
\begin{array}{l}{\left\langle\tau_{s}\right\rangle \sim \frac{\xi^{2 / 3}}{g_{s}}}, \\ {\langle\mathcal{V}\rangle \sim e^{a_{s}\tau_{s}}}.\end{array}
\end{equation}
Here $\tau_s$ is the small, heavy modulus, while the large light modulus is associated to the overall breathing mode corresponding to volume rescalings.

One can usefully obtain a 1-field effective potential for the light modulus by integrating out
the heavy modulus. In terms of the canonically normalized field $\varphi = \sqrt{\frac{2}{3}} \ln \mathcal{V}$, the resulting potential is
\begin{equation}\label{eq:pot}
V = e^{- \frac{\lambda \varphi}{M_P}} \big( A' - \varphi^{3/2}\big),
\end{equation}
for a constant $A'$, where in LVS the coefficient $\lambda = \sqrt{\frac{27}{2}}$. This value of $\lambda$ corresponds to the overall $\mc{V}^{-3}$ scaling of the potential and so is fixed in LVS. There is a single minimum located at a critical value $\varphi_c = \langle \phi \rangle$ of the field satisfying
\begin{equation}
 \frac{3}{2} \lambda\varphi_c^{\frac{3}{2}}- \varphi_c^{\frac{1}{2}} = \lambda A'.
\end{equation}
This corresponds to $\mc{V} \gg 1$ since $\varphi$ is logarithmic in the volume and $A'$ takes typical $\mc{O}(1 - 10)$ values. It is worth emphasising that as $\ln \mc{V} \sim \frac{1}{g_s}$, and there is a large discretuum of flux choices to fix $g_s$ to small values, in practice $\langle \mc{V} \rangle$ can be made as large as one wishes and is effectively a free parameter.
The corresponding vacuum realizes an AdS solution and one can verify that
\begin{equation}
V''(\varphi_c) = - \lambda^2 V_{min} \left(1-\frac{1}{2 \lambda \varphi_c} \right).
\end{equation}
Using $V_{min} \equiv -\frac{3 M_P^2}{R^2_{AdS}}$, this can be rewritten \cite{Conlon:2018vov}
\begin{equation}
V''(\varphi_c) = \frac{3 \lambda^2}{R_{AdS}^2} \big( 1 + \mathcal{O}(\ln \mathcal{V}^{-1}) \big).
\end{equation}
From a holographic perspective, this equation (combined with Eq. (\ref{eq:dim})) carries a striking implication: the conformal dimension of the volume modulus is fixed in the infinite volume limit, approaching in the $\mathcal{V} \to \infty$ limit  a universal value (with subleading corrections in $(\ln \mathcal{V})^{-1}$
\begin{equation}
\Delta_{\varphi} = \frac{3 \Big(1+ \sqrt{1+ \frac{4}{3}\lambda^2} \Big)}{2}.
\end{equation}
Moreover, a similar phenomenon holds for the interaction terms; repeated differentiations of Eq. (\ref{eq:pot}) give
\begin{equation}
\label{eq:phi}
V^{(n)}(\varphi)= (-\lambda)^n \frac{3 M_P^2}{R^2_{AdS}} \frac{n-1}{n!} \Big( \frac{\varphi}{M_P} \Big)^n \Big( 1+ \mathcal{O}\Big(\frac{1}{\lambda \varphi_c}\Big) \Big),
\end{equation}
whose form is now independent of specific details of the UV theory (such as $W_0$, the details of the Calabi-Yau or the choice of fluxes).
The volume axion has a shift symmetry and so its potential vanishes, leaving it massless, with the only interactions arising from its kinetic term
\begin{equation}
\label{eq:aaphi}
\mathcal{L}_{aa \varphi^{n-2}} = \Bigg(-\sqrt{\frac{8}{3}}\,\Bigg)^{n-2} \frac{1}{2(n-2)!}  \Big( \frac{\varphi}{M_P} \Big)^n.
\end{equation}
Furthermore, the structure of Eqs. (\ref{eq:phi}) and (\ref{eq:aaphi}) is radiatively stable, since quantum corrections give sub-leading effects in the large volume expansion \cite{Conlon:2018vov} and so leave the basic form of the potential of Eq. (\ref{eq:pot}) unaltered.

\subsection{CFT interpretation}

Given the uniqueness of the spectrum and interactions of low conformal dimension operators in the infinite volume limit, it is natural to view the low-dimension sector of LVS vacua as a small perturbation about this unique theory, which thus acquires a special status. In particular, the low dimensional sector of the putative CFT dual is characterized by a sparse spectrum, as the only single trace operators which retain a finite conformal dimension in the large volume limit are (as above) the volume modulus and its axion, plus the stress energy tensor. Crucially, the dimensions of the other moduli diverge as $\mathcal{V}$ is taken to infinity, as shown in Table \ref{tab:heavy}. Therefore, this sector of the theory is fully specified by the conformal dimensions of Table \ref{tab:model} and the interactions of (\ref{eq:phi}) - (\ref{eq:aaphi}).
\begin{table}[h!]
\centering
\begin{tabular}{|c|c|c|}
\hline
 Mode & Mass  & $\Delta$  \\
\hline
$\tau_s$ & $\frac{M_P \ln \mathcal{V}}{\mathcal{V}}$    &  $\mathcal{V}^{\frac{1}{2}}\ln \mathcal{V}$    \\
$T,U$ & $\frac{M_P }{\mathcal{V}}$    &  $\mathcal{V}^{\frac{1}{2}}$    \\
$\psi_{3/2}$ & $\frac{M_P W_0}{\mathcal{V}}$    &  $\mathcal{V}^{\frac{1}{2}}$ \\
\hline
\end{tabular}
\caption{Large $\mathcal{V}$ asymptotics of the masses and conformal dimensions for the single trace operators dual to
the other moduli in the minimal realization of LVS, specifically the complex structure moduli, the small Kähler modulus and the gravitino. A similar decoupling occurs for other modes like brane moduli. The KK and string modes are even heavier with $\Delta_{KK} \sim \mc{V}^{5/6}$ and $\Delta_{string} \sim \mc{V}$, while black hole states start occurring at $\Delta_{BH} \sim \mc{V}^{3/2}$.}
\label{tab:heavy}
\end{table}
\begin{table}[h!]
\centering
\begin{tabular}{|c|c|c|c|}
\hline
 Operator & Spin & Parity & $\Delta$  \\
\hline
$a$ & $0$    &  $-$  & 3  \\
$\varphi$ & $0$    &  $+$  & $\frac{3}{2}(1+\sqrt{19})$  \\
$T_{\mu \nu}$ & $2$    &  $+$  & 2  \\
\hline
\end{tabular}
\caption{Low dimensional single-trace operators in the spectrum of the CFT dual to LVS in the $\mathcal{V} \to \infty$ limit. Table taken from \cite{Conlon:2018vov}. }
\label{tab:model}
\end{table}
The latter can then be recast into $\mathcal{O}(\frac{1}{N^2})$ CFT data (OPE coeffiecients and anomalous dimensions) through the AdS/CFT correspondence by computing correlation functions.

From a holographic perspective, the appeal of LVS lies precisely in the fact that such computations become tractable due to the small number of states propagating inside Witten diagrams. It therefore represents both a serious scenario of moduli stabilisation and also an ideal playground to study the questions outlined in the introduction, namely whether Swampland constraints in AdS can be rephrased or understood in terms of the corresponding CFT.

\subsection{Connection to the Swampland}\label{ssc:swc}

The first observation is that some alterations of Eqs. (\ref{eq:phi}) - (\ref{eq:aaphi}) are clearly inconsistent from a stringy point of view, even if they may not appear fatal within the effective theory. Perhaps the simplest such class of modifications comprises changes in the structure of the axion kinetic term, modifying the large volume behaviour of the axion decay constant. With the original kinetic term
\begin{equation}\label{eq:kin}
\mathcal{L} \supset e^{-\sqrt{\frac{8}{3}} \frac{\varphi}{M_P}} \partial_{\mu}a \partial^{\mu}a,
\end{equation}
the axion decay constant scales as $f_a \sim M_P \mathcal{V}^{-\frac{2}{3}}$ and
vanishes in the $\mc{V} \to \infty$ limit. However, were $f_a$ to instead remain constant or even diverge, this would be in stark contrast with the evidence for sub-Planckian decay constants within quantum gravity as well as known behaviour in string theory. Indeed, explicit examples in String Theory show control issues systematically occurring when $f_a \gg M_P$ (e.g see \cite{Banks:2003sx,Conlon:2011qp,Baume:2016psm,Valenzuela:2016yny}).

For example, according to the axionic version of the Weak Gravity Conjecture \cite{ArkaniHamed:2006dz}, $f_a$ should satisfy the inequality
\begin{equation}\label{eq:weakax}
f_a S \lesssim M_P,
\end{equation}
where $S$ is the action of the leading instanton depending on the axion
%%contributing to the axion potential, written as
%%\begin{equation}
%%V(a)= \Lambda^4 \sum_q e^{-qS} (1-\cos (qa)),
%%\end{equation}
and $S$ is required to be smaller than one in order for the single instanton approximation to be valid.
Here, the action of a bulk D3-instanton is $S \sim \mc{V}^{2/3}$ and so Eq. (\ref{eq:weakax}) would be violated by $f_a \sim M_P$.
%%If $f_a \gg M_P$, \ref{eq:weakax} implies higher instanton contributions have to be taken into account, and this has two effects. A first point, of phenomenological relevance, is that although the periodicity $2 \pi f_a$ grows, the size of the regions where the potential is monotonic remains roughly the same, due to the sum of various oscillating contributions with shorter periods. More importantly, the energy density of the potential approaches $\Lambda^4$, so that the description is at danger of escaping its regime of validity.

This inconsistent behaviour can, for example, be achieved by switching the sign of the exponential in Eq. (\ref{eq:kin}) -- corresponding to $f_a \sim M_P \mathcal{V}^{\frac{2}{3}}$ -- or by including only a finite number of terms in its expansion. Such statements can now be translated into statements on the structure of the EFT Lagrangian; the above sign inversion, for instance, corresponds to an additional factor of $(-1)^n$ in Eq.(\ref{eq:aaphi}). More generally, when expanding about any vacuum the sign of the linear coupling $\frac{\varphi}{M_P} \partial_{\mu} a \partial^{\mu} a$ is clearly equivalent to the sign of $\frac{\partial f_a(\varphi)}{\partial \varphi}$, so any requirement that $\frac{\partial f_a}{\partial \varphi} < 0$ in the asymptotic regime is equivalent to a negative sign for the latter coupling as well.

As another example, one might wonder whether potentials resembling LVS -- but with a generic $\lambda$ -- can be realized in string compactifications. One reason to suppose this is \emph{not} possible is that, if $\lambda \leq \sqrt{6} $, the dependence on the volume is such that the potential grows parametrically faster than the string scale $M_s^4$, so $V/M_s^4 \to \infty$ as $\mc{V} \to \infty$ -- a behaviour that looks problematic in a limit of weak coupling and large volumes. A value of $\lambda = \sqrt{6}$ corresponds to
\begin{equation}
V \propto \frac{M_P^4}{\mathcal{V}^2} \sim M_S^4
\end{equation}
since $M_S \sim \frac{M_P}{\sqrt{\mathcal{V}}}$.

In the other direction, for the case of $\lambda \gg 1$ the leading contribution to the potential would scale with a very high power of the volume. This also appears unstable against quantum loop corrections which would be expected to re-introduce a scaling with a lower power of the volume \cite{Cicoli:2007xp, Burgess:2010sy}. One may therefore hope that all values of $\lambda$ outside  a finite interval should lie in the Swampland, or even adopt the extreme point of view that only $\lambda = \sqrt{\frac{27}{2}}$ is allowed.

The above arguments then provide a motivation for a holographic approach.
Within this holographic approach, there are two basic questions to be asked:
\begin{itemize}
\item Can one turn swampland constraints on the AdS Lagrangian into well-defined statements about CFT properties?
\item Do swampland modifications to the AdS side clearly translate into violations of fundamental properties of the putative CFT dual?
\end{itemize}

Anticipating subsequent results, we can say that, although the first question seems to admit a positive response, the second question seems harder to answer. In particular, we will show how the above Swampland conditions translate into a statement on CFT anomalous dimensions which is surprisingly similar to known causality and unitarity constraints, but is nonetheless not known to be valid in general.

\section{Holographic CFTs and Consistency Conditions}\label{sc:boot}

We now turn to a general discussion of holographic CFTs, and the form of constraints on them that can exist, before returning to applications to LVS.

\subsection{Holographic CFTs}

According to the AdS/CFT correspondence, correlation functions in a conformal field theory can be calculated on the gravity side by evaluating boundary correlators involving the corresponding fields in AdS. If the gravity dual is weakly coupled, such computations can be carried out explicitly in perturbation theory, and the CFT is said to be holographic. From the point of view of string theory, this is the case when $R_{AdS} \gg \ell_S$ and $g_s \ll 1$, so quantum corrections (both in spacetime and on the worldsheet) can be neglected. According to the AdS/CFT dictionary the 't Hooft coupling $\lambda = g^2 N$ is mapped to $\lambda \sim (R/\ell_S)^d$ and so must be much larger than one in this limit,
implying that holographic CFTs are always characterized by $N \gg 1$, or an equivalent expansion parameter when the theory does not admit an explicit large-$N$ description. With a slight abuse of notation (which conforms to what has become standard use in the literature), we shall always denote this expansion parameter as $1/N$, and operators will be referred to as single or double trace according to the scaling of their two-point function with $N$. In the LVS minimum, $R_{AdS} \sim \mc{V}^{3/2}$, and so the stabilised volume plays the role of the large $N$ parameter.

A second peculiarity of holographic CFTs is that they are characterized by a large gap in the spectrum of conformal dimensions, which are related to the AdS mass through the formula
\begin{equation}
\Delta (\Delta -d) = m^2 R^2_{AdS}
\end{equation}
already cited in the introduction. In holographic CFTs there is a large $\Delta_{Gap} \gg 1$ between the low-lying scalar, vector and graviton modes to higher-spin operators. For example, all the heavy stringy modes have $m \sim 1/\ell_S \gg 1/R_{AdS}$ and correspond to fields with large conformal dimensions in the weak coupling regime.

It is interesting to note that the converse of this statement has been conjectured to hold, i.e. that all large-$N$ CFTs with a gap should be characterized by a weakly coupled, local bulk dual. In \cite{Heemskerk:2009pn}, this was proven at order $1/N^2$ through a bijective mapping of solutions to the Bootstrap crossing equations to bulk interaction vertices for scalars. Incidentally, this shows that the requirement of crossing symmetry alone is not enough to constrain the form of the low energy Lagrangians mentioned at the beginning.

From a practical point of view, the spectrum of holographic CFTs at the first non-trivial order in $1/N^2$ includes a finite number of single-trace primaries $\mathcal{O}_1, \mc{O}_2, \ldots \mathcal{O}_m $ plus double-trace (and higher-trace) operators, which are built from any two (or more) of the $\mathcal{O}_i$. For scalars, the double-trace operators are schematically of the form
\begin{equation}
\mathcal{O}_i \Box^n \partial_{\mu_1} ... \partial_{\mu_\ell} \mathcal{O}_j,
\end{equation}
and there is a single double-trace primary for each value of $n$ and $\ell$, denoted as $[\mathcal{O}_i \mathcal{O}_j]_{n,\ell}$. Their conformal dimension can be written as
\begin{equation}
\Delta_{n,\ell} = \Delta_i + \Delta_j + 2n + \ell + \gamma(n,\ell),
\end{equation}
where the classical contribution $\Delta_i + \Delta_j + 2n + \ell$ can be split from the smaller
anomalous dimension $\gamma(n,\ell)$ that is $\mc{O}(1/N^2)$. From an AdS perspective, the double trace operator corresponds to a 2-particle state in AdS and the anomalous dimension corresponds to its binding energy. Of course, in a full-fledged CFT scalars are not the whole story, and there will also be the stress energy tensor plus the operators built by combining the stress tensor with other primaries.

\subsection{The Bootstrap}

The bootstrap program, originally dating to the early 1970s \cite{Ferrara:1973yt,Polyakov:1974gs,Mack:1975jr}, rests on the philosophy that conformal symmetry alone is powerful and restrictive enough to impose significant constraints on the space of allowed theories, allowing interesting results to be derived with little external input other than the CFT axioms. On the technical side, a fundamental property is the existence of a convergent OPE expansion
\begin{equation}\label{eq:ope1}
\mathcal{O}_1(x_1) \mathcal{O}_2(x_2) = \sum_{k \,\, \text{primary}} f_{12}^k C(x_{12}, \partial_{12}) \mathcal{O}_k(x_2),
\end{equation}
where the operator $C(x, \partial)$ is fully determined by conformal symmetry, and $f_{12k}$ is the only arbitrary coefficient appearing in the $\langle \mathcal{O}_1 \mathcal{O}_2 \mathcal{O}_k \rangle$ three-point function,
namely
\begin{equation}\label{eq:3f}
\langle \mathcal{O}_1(x_1) \mathcal{O}_2(x_2) \mathcal{O}_k(x_3) \rangle= \frac{f_{12k}}{|x_{12}|^{\Delta_1+\Delta_2-\Delta_k} |x_{23}|^{\Delta_k+\Delta_2-\Delta_1}|x_{13}|^{\Delta_1+\Delta_k-\Delta_2}}.
\end{equation}

When inserted inside a 4-point correlator, this generates an expansion in a universal basis of functions, the conformal blocks.
For identical operators,
\begin{equation}
\langle \mathcal{O}(x_1) \mathcal{O}(x_2)  \mathcal{O}(x_3) \mathcal{O}(x_4) \rangle =  \frac{1+\sum_{\Delta, \ell} \big(f^{\Delta,\ell}_{\mathcal{O} \mathcal{O}}\big)^2 G_{\Delta, \ell} (u,v)}{|x_{12}|^{2 \Delta_1} |x_{34}|^{2 \Delta_3} } \equiv
\frac{1+ \mathcal{A}(u,v)}{|x_{12}|^{2 \Delta_1} |x_{34}|^{2 \Delta_3} },
\end{equation}
where $u$ and $v$ the conformal cross ratios, defined as
\begin{equation}
u = \frac{x_{12}^2 x_{34}^2}{x_{13}^2 x_{24}^2},  \quad \quad \quad  v = \frac{x_{14}^2 x_{23}^2}{x_{13}^2 x_{24}^2}.
\end{equation}
Although they are completely fixed by conformal invariance, the explicit form of the conformal blocks is only known in even dimension, where they satisfy a separable differential equation.

Thus, all the dynamical information about the CFT is contained in the spectrum of the theory, given by the pairings ${\Delta_i,\ell_i}$, and the OPE coefficients $f_{12}^k$.
In particular, with a 4-point function one can perform the contraction (\ref{eq:ope1}) in two inequivalent ways, to obtain the celebrated Bootstrap equation
\begin{equation}\label{eq:cross}
\sum_{\Delta, \ell} \big(f^{\Delta,\ell}_{\mathcal{O} \mathcal{O}}\big)^2 \left(\frac{v^{\Delta_{\phi}} G_{\Delta, \ell}(u, v)-u^{\Delta_{\phi}} G_{\Delta, \ell}(v, u)}{u^{\Delta_{\phi}}-v^{\Delta_{\phi}}}\right) = 1.
\end{equation}

In principle, with a holographic CFT it is possible to calculate any correlator with Witten diagrams in AdS, and then expand into conformal blocks to recover OPE coefficients and anomalous dimensions. However, this is hardly ever achievable in practice -- computations are extremely long even for the simplest of diagrams and 3d conformal block are not known in a closed form. Furthermore, the full correlators often contain much more information than is needed, and one is left wondering whether it would be possible to skip some of the intermediate steps.

For this reason, bootstrap techniques often yield powerful insights, both in terms of uncovering general properties and also performing specific calculations. In a variety of examples, the crossing equation (\ref{eq:cross}) was used to recover the leading order OPE data in large-N theories, corresponding to both contact \cite{Heemskerk:2009pn} and exchange \cite{Alday:2017gde} diagrams in AdS. In this context, a general formula for the special case $n=0$ was worked out in \cite{Costa:2014kfa} with the Mellin amplitude formalism, to be introduced below. This will be discussed extensively in section \ref{ssc:form}, and will form the basis of our analysis.

\subsection{Mellin Amplitudes}

The S-matrix does not exist in AdS as any definition of physical asymptotic states is troublesome; the only real observables are correlators or functions thereof. Heuristically, this arises because AdS behaves like a system of ``particles in a box'', where interactions between different constituents cannot be switched off and there are no `states at infinity'.

However, there does exist
a representation of AdS correlators which exhibits striking similarities to flat space scattering amplitudes,
and can be used to derive the latter in certain limits where the radius of AdS goes to infinity. These are the Mellin amplitudes.

Starting from an $n$-point correlator, the Mellin transform is defined as
\begin{equation}
A\left(x_{i}\right) \supset \left\langle\mathcal{O}_{1}\left(x_{1}\right) ...\ldots \mathcal{O}_{n}\left(x_{4}\right)\right\rangle_{\mathrm{c}}=\prod_{1 \leq i<j \leq n} \int_{-i \infty}^{+i \infty} \frac{\text{d} \delta_{ij}}{2 \pi i} \, M(\delta_{i j})  \Gamma\left(\delta_{i j}\right)\left(x_{i j}^{2}\right)^{-\delta_{i j}},
\label{mellintransform}
\end{equation}
with the quantity $M(\delta_{ij})$ known as the Mellin amplitude \cite{Mack:2009mi,Mack:2009gy,Penedones:2010ue}.
In Eq. (\ref{mellintransform}) the integrations should be carried out in such a way that the poles arising from the Mellin amplitude or a given gamma function all be placed on one side of the contour. Defining $\delta_{ii}= -\Delta_i$, one can show that conformal invariance implies
\begin{equation}
\sum_i \delta_{ij} = 0.
\end{equation}
After imposing these constraints, there are $n(n-3)/2$ independent variables $\delta_{ij}$ for an $n$-point function, which is the same as the number of kinematic invariants for an $n$-particle amplitude.\footnote{In both cases, this counting is modified for $n > d+2$.}
It is then possible to introduce fictitious variables $p_i$ that satisfy
\begin{equation}
p_i \cdot p_j = \delta_{ij}, \quad \quad \quad \sum_{i=1}^n p_i = 0,
\end{equation}
representing the analogue of momentum conservation for conformal correlators. For 4-point correlators, it is then natural to use Mandelstam-like variables\footnote{Here the definition of $t$ has a constant shift with respect to the canonical one only to simplify some formulas involving Mack polynomials.}
\begin{equation}
\begin{split}
s= -(p_1+p_2)^2 = \Delta_1 + \Delta_2-2 \delta_{12}, \quad & \quad  u= -(p_1+p_4)^2 = \Delta_1 + \Delta_4 -2 \delta_{14}, \\
t= -(p_1+p_3)^2-\Delta_1 &-\Delta_4 = \Delta_3-\Delta_4-2 \delta_{13},
\end{split}
\end{equation}
which obey the relation
\begin{equation}
s+t+u = \Delta_2 +\Delta_3.
\end{equation}
For a 4-point function, the reduced amplitude $\mc{A}(u, v)$ is defined as
\begin{equation}
A\left(x_{i}\right)=\frac{1}{\left(x_{12}^{2}\right)^{\frac{\Delta_{1}+\Delta_{2}}{2}}\left(x_{34}^{2}\right)^{\frac{\Delta_{3}+\Delta_{4}}{2}}}\left(\frac{x_{24}^{2}}{x_{14}^{2}}\right)^{\frac{\Delta_{1}-\Delta_{2}}{2}}\left(\frac{x_{14}^{2}}{x_{13}^{2}}\right)^{\frac{\Delta_{3}-\Delta_{4}}{2}} \mathcal{A}(u, v),
\end{equation}
and it can be expressed as a function of the Mellin amplitude
\begin{equation}\label{eq:mel}
\begin{aligned} \mathcal{A}(u, v)=& \int_{-i \infty}^{i \infty} \frac{d t d s}{(4 \pi i)^{2}} M(s, t) u^{s / 2} v^{-(s+t) / 2} \Gamma\left(\frac{\Delta_{1}+\Delta_{2}-s}{2}\right) \Gamma\left(\frac{\Delta_{3}+\Delta_{4}-s}{2}\right) \\ & \Gamma\left(\frac{\Delta_{34}-t}{2}\right) \Gamma\left(\frac{-\Delta_{12}-t}{2}\right) \Gamma\left(\frac{t+s}{2}\right) \Gamma\left(\frac{t+s+\Delta_{12}-\Delta_{34}}{2}\right). \end{aligned}
\end{equation}

Further evidence for the analogy comes from the explicit computation of Mellin amplitudes, at least in the simple cases where it is possible to do so. For a scalar contact interaction between $n$ different fields
\begin{equation}
\mathcal{L}_{int} = g \varphi_1 \varphi_2... \varphi_n,
\end{equation}
the Mellin amplitude is a constant,
\begin{equation}
M(\delta_{ij})= \frac{g \pi^{\frac{d}{2}}}{2} \Gamma \Big( \frac{\sum \Delta_i-d}{2}\Big){\prod_{i=1}^n} \frac{1}{\Gamma(\Delta_i)}.
\end{equation}
If derivatives are added to the vertex, as in
\begin{equation}
\mathcal{L}_{int} = g \nabla ...\nabla\varphi_1 \nabla ...\nabla\varphi_2... \nabla ...\nabla\varphi_n,
\end{equation}
then the Mellin amplitude picks up powers of the fictional momenta corresponding to the number of derivatives acting on the field \cite{Penedones:2016voo}
\begin{equation}
M(\delta_{ij})= \frac{g \pi^{\frac{d}{2}}}{2} \Gamma \Big( \frac{\sum \Delta_i-d+2N}{2}\Big){\prod_{i=1}^n} \frac{1}{\Gamma(\Delta_i+\beta_i)} {\prod_{i<j}^n}(-2 \delta_{ij})^{\alpha_{ij}}+...,
\end{equation}
where $\alpha_{ij}$ is the number of derivatives acting on $\varphi_i$ and $\varphi_j$, $\beta_i = \sum_{j\neq i} \alpha_{ij}$ and $2N$ is the total number of derivatives.
Going one step further, a diagram describing the exchange of a bulk scalar dual to a single trace operator $\mathcal{O}$ of dimension $\Delta_{\mathcal{O}}$ and spin $\ell_{\mathcal{O}}$, say in $s$-channel, has the form \cite{Fitzpatrick:2011ia}
\begin{equation}
M(s,t) = f_{12\mathcal{O} } f_{\mathcal{O}34} \sum_m \frac{\mathcal{Q}_{\ell_{\mathcal{O}},m}(t)}{s-\Delta_{\mathcal{O}}+\ell_{\mathcal{O}}-2m},
\end{equation}
where the $\mathcal{Q}_{\ell_{\mathcal{O}},m}(s,t)$, known as Mack Polynomials, are completely determined by conformal symmetry.
Introducing the Pochhammer symbol
\begin{equation}
(a)_m= \frac{\Gamma(a+m)}{\Gamma(a)} = a(a+1)...(a+m-1),
\end{equation}
their explicit form can be conveniently parametrized by the new polynomials $Q_{J,m}(s)$ satisfying \cite{Costa:2012cb}
\begin{equation}
\begin{aligned}
\mathcal{Q}_{J, m}(s)=&-\frac{2 \Gamma(\Delta+J)(\Delta-1)_{J}}{4^{J} \Gamma\left(\frac{\Delta+J+\Delta_{12}}{2}\right) \Gamma\left(\frac{\Delta+J-\Delta_{12}}{2}\right) \Gamma\left(\frac{\Delta+J+\Delta_{34}}{2}\right) \Gamma\left(\frac{\Delta+J-\Delta_{34}}{2}\right)} \\
& \frac{Q_{J, m}(s)}{m !(\Delta-h+1)_{m} \Gamma\left(\frac{\Delta_{1}+\Delta_{2}-\Delta+J}{2}-m\right) \Gamma\left(\frac{\Delta_{3}+\Delta_{4}-\Delta+J}{2}-m\right)}.
\end{aligned}
\end{equation}
Imposing the normalization $Q_{J,m}(s)  = s^J+\mathcal{O}(s^{J-1})$, these are given by
\begin{equation}\label{eq:mackp}
Q_{J, 0}(s)=\frac{2^{J}\left(\frac{\Delta_{12}+\tau}{2}\right)_{J}\left(\frac{\Delta_{34}+\tau}{2}\right)_{J}}{(\tau+J-1)_{J}} {}_3F_{2}\left(-J, J+\tau-1, \frac{\Delta_{34}-s}{2} ; \frac{\tau+\Delta_{12}}{2}, \frac{\tau+\Delta_{34}}{2} ; 1\right).
\end{equation}

The appearance of poles in the Mandelstam-like variables corresponding to the exchange of operators of different spin is reminiscent of flat space scattering amplitudes. Quite remarkably, this is not only a feature of weakly coupled gravity duals, but instead a generic property valid for all conformal theories. Indeed, given the OPE expansion
\begin{equation}
\mathcal{O}_{1}\left(x_{1}\right) \mathcal{O}_{1}\left(x_{2}\right)=\sum_{k} f_{12 k}\left(x_{12}^{2}\right)^{\frac{\Delta_{k}-\Delta_{1}-\Delta_{2}}{2}}\left[\mathcal{O}_{k}\left(x_{2}\right)+c x_{12}^{2} \partial^{2} \mathcal{O}_{k}\left(x_{2}\right)+\ldots\right],
\end{equation}
in the OPE limit $x_{12}^2 \rightarrow 0$ one can perform the integral over $\delta_{12}$  by summing over the corresponding poles, denoted by a tilde.
\begin{equation}
\left\langle\mathcal{O}_{1}\left(x_{1}\right) \mathcal{O}_{1}\left(x_{2}\right) \ldots\right\rangle=\sum_{\tilde{\delta}_{12}}\left(x_{12}^{2}\right)^{-\tilde{\delta}_{12}} \prod_{\substack{1<i<j,\\ 2<j}} \int  \frac{\text{d} \delta_{ij}}{2 \pi i} \operatorname{Res}_{\tilde{\delta}_{12}} M \left(\delta_{i j}\right) \Gamma \left(\delta_{i j}\right)\left(x_{i j}^{2}\right)^{-\delta_{i j}},
\end{equation}
Matching terms in an expansion in inverse powers of $x_{12}^2$ it is then possible to recover poles at
\begin{equation}
s = \Delta_{\mathcal{O}}+2m, \quad \quad \quad m = 1,2 \ldots ,
\end{equation}
with a residual that is proportional to $f_{12 \mathcal{O}}$. A careful derivation of factorization formulas for Mellin amplitudes can be found in \cite{Goncalves:2014rfa}.

\subsubsection{Applications: a formula for $\gamma(0,\ell)$}\label{ssc:form}

As an application of this formalism, we now derive a tree-level formula to compute the $\gamma(0,\ell)$ anomalous dimensions of a holographic CFT in terms of the associated Mellin amplitude, used for the first time in \cite{Costa:2014kfa} for the study of gravitational exchange diagrams.

Given two single trace primaries $\mathcal{O}_1,\mathcal{O}_3$ of dimensions $\Delta_1$ and $\Delta_3$, the anomalous dimensions of the double trace operators $[\mathcal{O}_1\mathcal{O}_3]_{0,\ell}$ are given by
\begin{equation}\label{eq:adf}
\begin{split}
\gamma(0,\ell) = - \int_{-i \infty}^{+ i \infty} & \frac{ds}{2 \pi i } M(s,0)\,\, _3F_2(-\ell, \Delta_1+\Delta_3 + \ell-1,\frac{s}{2}; \Delta_1, \Delta_3;1 ) \\ & \times \Gamma \Big(\Delta_1-\frac{s}{2}\Big)\, \Gamma \Big(\Delta_3-\frac{s}{2} \Big)\, \Gamma \Big(\frac{s}{2}\Big)^2,
\end{split}
\end{equation}
where $M(s,t)$ is the Mellin amplitude corresponding to the correlator
\begin{equation}\label{eq:cor}
\langle \mathcal{O}_{1}(x_1) \mathcal{O}_{1}(x_2) \mathcal{O}_3(x_3)  \mathcal{O}_3(x_4) \rangle.
\end{equation}
This expression only includes the analytic contribution in the spin arising from exchange diagrams, and is not sensitive to the presence of contact terms.\footnote{See \ref{ssc:ma}.}

To prove it, we compare the conformal block decomposition of the correlator with an appropriate expansion of the Mellin amplitude.
As the double trace operators above are only contained in the OPE of $\mathcal{O}_1$ with $\mathcal{O}_3$, the conformal block decomposition contains $\gamma(0,\ell)$ at order $\frac{1}{N^2}$ only in the mixed channels. If we denote by $\mathcal{A}'(u,v)$ the stripped correlator associated to
$\langle \mathcal{O}_3(x_1)
 \mathcal{O}_{1}(x_2)   \mathcal{O}_{1}(x_3) \mathcal{O}_3(x_4) \rangle$,
then
\begin{equation}
\mathcal{A}(u,v)v^{\frac{\Delta_1+\Delta_3}{2}} =\mathcal{A}'(v,u) u^{2 \Delta_1},
\end{equation}
and we can decompose $\mathcal{A}'(v,u)$ in s-channel conformal blocks.
\begin{equation}
\mathcal{A}(u,v) \supset \Big(\frac{u}{v} \Big)^{\frac{\Delta_1+\Delta_3}{2}}
u^{\frac{\Delta_1-\Delta_3}{2}}\sum_{\Delta, \ell} \big( f_{[13]_{0,\ell}}^{1,3} \big)^2 G(v,u)_{\Delta_1+\Delta_3+2n+\ell+\frac{\gamma(n,\ell)}{N^2}, \ell}^{\Delta_{31},\Delta_{13}}
\end{equation}
where the $f_{[13]_{0,\ell}}^{1,3}$ are OPE coefficients of $[\mathcal{O}_1\mathcal{O}_3]_{0,\ell}$ with $\mathcal{O}_1$ and $\mathcal{O}_3$. To isolate the contribution of $\gamma(0,\ell)$,
\begin{equation}
G(v,u)_{\Delta_1+\Delta_3+2n+\ell+\frac{\gamma(n,\ell)}{N^2}, \ell}^{\Delta_{31},\Delta_{13}} =
G(v,u)_{\Delta_1+\Delta_3+2n+\ell, \ell}^{\Delta_{31},\Delta_{13}}+ \frac{\gamma(n,\ell)}{N^2} \partial_{\Delta} G_{\Delta_1+\Delta_3+2n+\ell, \ell}^{\Delta_{31},\Delta_{13}}.
\end{equation}
Since conformal blocks can always be expanded as
\begin{equation}
G^{ \{ \Delta_i \} }_{\Delta, \ell}(u,v)= u^{\frac{\Delta-\ell}{2}} \sum_{m=0}^{\infty} g^{ \Delta
 ,\ell}_{\{ \Delta_i \},m}(v)u^m,
\end{equation}
the derivatives translate into the appearance of logarithms which, by analyticity of the conformal blocks, cannot be generated anywhere else in the sum. Therefore, the only log-singular term multiplying the lowest power of $u$ is
\begin{equation}\label{eq:cbex}
\mathcal{A}(u,v) \supset \frac{ u^{ \Delta_1} \log (v)}{2}
\sum_{\ell} \gamma(0,\ell) \big( f_{[13]_{0,\ell}}^{1,3} \big)^2 g^{ \Delta
 ,\ell}_{\{ \Delta_i \},0} (u).
\end{equation}
At this point it is convenient to introduce an integral representation \cite{Costa:2012cb} for $g_0$, in terms of the modified Mack polynomials defined in (\ref{eq:mackp}) :
\begin{equation}
g^{ \Delta
 ,\ell}_{\{ \Delta_i \},0} (u)=\alpha(\Delta_i,\ell) u^{- \Delta_1} \int_{-i \infty}^{+i \infty} \frac{ds}{8 \pi i} u^{-\frac{s}{2}} Q_{J,0}(s-\Delta_1+\Delta_3)  \Gamma \big(-\frac{s}{2}\big)^2 \Gamma\big(\Delta_1+\frac{s}{2}\big) \Gamma \left(\Delta_3+\frac{s}{2}\right).
\end{equation}
Inserting this last equation into (\ref{eq:cbex}) results in an expression that is highly reminiscent of Mellin amplitudes, to which we now turn our attention.

In fact, the same quantity can be rewritten as
\begin{equation}
 \mathcal{A}(u, v)= \int_{-i \infty}^{i \infty} \frac{d t d s}{(4 \pi i)^{2}} M(s, t) u^{s/2} v^{t / 2} \Gamma \big(\Delta_1-\frac{s}{2}\big)  \Gamma\big(\Delta_3-\frac{s}{2}\big) \Gamma \big(\frac{s+t}{2}\big)^2 \Gamma \big(-\frac{t}{2}\big)^2,
\end{equation}
where we have used the reflection property $M(s,t) = M(s,-s-t)$ \footnote{A consequence of $1 \leftrightarrow 2 / 3 \leftrightarrow 4$ symmetry.} to switch integration variables with respect to (\ref{eq:mel}). Since the OPE limit in the crossed channel corresponds to $v \rightarrow 0$, one can integrate with respect to $t$ by closing the contour on the right-hand plane. On the positive axis, there are double poles located at integer values of $t=2n$.
Using
\begin{equation}
\text{Res} \Big( \Gamma \big(-\frac{t}{2}\big)^2 f(t) \Big)_{t=2n} = \frac{4}{(n!)^2}f'(2n) + \frac{2 (-1)^n C_n }{n!} f(2n),
\end{equation}
where the form of the coefficients $C_n$ is irrelevant for our purposes. We can then isolate the logarithmic contribution
\begin{equation}\label{eq:melf}
\mathcal{A}(u, v)= \frac{\log v}{2} \sum_{n=0}^{\infty} \frac{v^n}{(n!)^2}\int_{-i \infty}^{+i \infty} \frac{d s}{2 \pi i} M(s, 2n) u^{s/ 2} \Gamma \big(\Delta_1-\frac{s}{2}\big)  \Gamma\big(\Delta_3-\frac{s}{2}\big) \Gamma \left(\frac{s+2n}{2}\right)^2 .
\end{equation}

Since the integrals in (\ref{eq:cbex}) and (\ref{eq:melf}) are equal for any value of $u$, we can equate the integrands\footnote{Up to terms which are not compatible with the structure of a Mellin amplitude $M(s,0)$ in a large-N theory with a weakly coupled dual.} to obtain
\begin{equation}\label{eq:conf}
\sum_{\ell=0}^{\infty} \gamma(0,\ell) \big( f_{[13]_{0,\ell}}^{1,3} \big)^2
Q_{\ell,0}(-s+\Delta_1-\Delta_3) \alpha(\Delta_i,\ell) = -4 M(s,0).
\end{equation}
 Incidentally, in the case of a polynomial amplitude (dual to contact vertices on AdS) Eq. (\ref{eq:conf}) becomes a finite dimensional linear system for the $\gamma(0,\ell)$, correctly reproducing some results contained in \cite{Heemskerk:2009pn} (and not captured by (\ref{eq:adf})). Finally, it is possible to isolate the individual $\gamma(0,\ell)$ through a projection on the corresponding Mack polynomials, by integrating (\ref{eq:conf}) with respect to the measure in (\ref{eq:melf}). This is because they satisfy the orthogonality relation \cite{Costa:2012cb}
\begin{equation}
\begin{aligned}  & \int_{-i \infty}^{i \infty} \frac{d s}{4 \pi i} Q_{J, 0}(s) Q_{J^{\prime}, 0}(s)  \Gamma\left(\frac{\Delta_{34}-s}{2}\right) \Gamma\left(\frac{-\Delta_{12}-s}{2}\right) \Gamma\left(\frac{\tau+s}{2}\right) \times  \\ \times
& \Gamma\left(\frac{\tau+s+\Delta_{12}-\Delta_{34}}{2}\right) =
 \delta_{J, J^{\prime}} \frac{(-4)^J \Gamma \big(\Delta_1+J\big)^2
\Gamma \big(\Delta_3+J \big)^2 J!}{\Gamma(\Delta_1+\Delta_3+2J) (\Delta_1+\Delta_3+J-1)_J},
\end{aligned}
\end{equation}
with the choice of normalization in (\ref{eq:mackp})
The result then follows from the generalized free theory coefficients in \cite{Fitzpatrick:2011dm}
\begin{equation}
\big( f_{[13]_{0,J}}^{1,3} \big)^2 = \frac{(-1)^{J} (\Delta_1)_{J} (\Delta_3)_{J}}{(\Delta_1+\Delta_3+J-1)_{J} J !}
\end{equation}
and the normalization factor
\begin{equation}
\alpha(\Delta_i,J)= \frac{2 \Gamma(\Delta_1+\Delta_3+2J)(\Delta_1+\Delta_3+J-1)_{J}}{4^{J} \Gamma \big(\Delta_1+J\big)^2
\Gamma \big(\Delta_3+J\big)^2}.
\end{equation}

\subsection{Positivity bounds}

In the last section, we saw that sign changes in the effective field theory Lagrangians of string vacua are able to move a theory in or out of the swampland. This is reminiscent of positivity bounds in QFTs deriving from unitarity and analyticity. For our purposes, we will mostly be concerned with two specific realization of this idea:
\begin{itemize}
\item S-Matrix positivity bounds, derived for the first time in \cite{Adams:2006sv}. These crucially rely on the assumption of analyticity\footnote{excluding branch cuts on the real axis and isolated singularities, as is usual in physics.} in the kinematic variables, an idea that is closely related to causality.
\item Constraints on the sign of the anomalous dimensions in Lorentzian CFTs, such as those derived in \cite{Komargodski:2012ek, Hartman:2015lfa, Kundu:2020gkz}.
\end{itemize}
These results often trace back to an application of the optical theorem, which relates a generic amplitude to an integral over intermediate states. In its most general form, it can be stated as \cite{Schwartz:2013pla}
\begin{equation}\label{eq:opt}
\mathcal{A}(i \rightarrow f )-  \mathcal{A}^*(f \rightarrow i ) = (2 \pi)^4 \delta^{(4)}(\sum_{j \in I} p_j - P_X)
\sum_X \int \text{d} \Pi_X   \mathcal{A}(i \rightarrow X ) \mathcal{A}^*(X \rightarrow f ).
\end{equation}
In the case of elastic scattering (i.e when the inital and final states coincide), the LHS of (\ref{eq:opt}) reduces to the imaginary part of the amplitude, while the RHS becomes manifestly positive: it is this sign that acts as a source for the positivity bounds. If one further specializes to two particle states, the RHS becomes the total cross section for the process under consideration.

\subsubsection{Positivity in CFTs}\label{sssc:pos}
In Conformal Field Theories, the idea of scattering is ill-defined, and so it is not straightforward to make contact with the usual implications of unitarity stemming from the Optical Theorem. However, an ingenious way to circumvent the problem (first devised in \cite{Komargodski:2012ek,Fitzpatrick:2012yx}) is to modify a CFT by a relevant perturbation that causes it to flow to a gapped phase in the infrared. In particular, the authors considered elastic scattering between the lowest-mass state in the theory, which now has a mass gap, and an arbitrary external operator $\mathcal{O}$ through a space-like virtual exchange. This procedure is in close analogy with Deep Inelastic Scattering (DIS) processes used to study the structure of hadrons, by bombarding them with energetic light leptons, and can thus be referred to as a DIS gedanken experiment.

With the additional assumption of polynomial boundedness on off-shell amplitudes\footnote{Which is not guaranteed by the Froissart bound.} and exploiting the positivity of the total cross section, it was shown in \cite{Komargodski:2012ek} that the minimal twist operators appearing in the OPE of $\mathcal{O}$ with its adjoint have to obey the convexity property
\begin{equation}\label{eq:pos}
\frac{\tau^*(\ell_3)-\tau^*(\ell_1)}{\ell_3-\ell_1} \leq \frac{\tau^*(\ell_2)-\tau^*(\ell_1)}{\ell_2-\ell_1}.
\end{equation}
Here the twist of an operator is defined as $\tau = \Delta - \ell$, and the relevance lies in the fact that the Lorentzian OPE is dominated in the OPE limit by low twist operators. More precisely, the $\tau^*(\ell)$ in (\ref{eq:pos}) is the twist of the lowest dimension operator of spin $\ell$ appearing in the OPE of a given $\mathcal{O}$ with itself, and as such is called a minimal twist operator. This property holds only if $\ell$ is higher than a certain critical spin $\ell \geq \ell_c$, which is determined by the leading exponent in the (polynomial) high energy limit of some amplitude, and is therefore not calculable a priori within this approach. However, in \cite{Costa:2017twz}, the authors used the Lorentzian OPE inversion formula of \cite{Caron-Huot:2017vep} to extend these relations for any continuous value of the spin $\ell >1$, without having to resort to any of the assumptions cited above.

The lower bound is in accordance with practical examples, where the theorem is found to hold for $\ell_c \geq 2$, and usually satisfied by the stress energy tensor -- which has the lowest twist possibly allowed by unitarity, $\tau= d-2$. More generally, the minimal twist operators that saturate the bounds for an arbitrary spin $\ell$ are appropriate combinations involving stress energy tensors. A notable exception is when gravitational interactions are very suppressed with respect to other couplings in the theory -- for example if the AdS dual contains interactions suppressed by a scale $\Lambda \ll M_P$ and gravity can be integrated out. In a holographic CFT of this kind, one can turn (\ref{eq:pos}) into
\begin{equation}\label{eq:pos1}
\frac{\gamma(0,\ell_3)-\gamma(0,\ell_1)}{\ell_3-\ell_1} \leq \frac{\gamma(0,\ell_2)-\gamma(0,\ell_1)}{\ell_2-\ell_1},
\end{equation}
where $\gamma(n,\ell)$ is the anomalous dimension of the double trace operators made out of the lowest dimension scalar in the theory. An immediate consequence of the convexity of Eq. (\ref{eq:pos1}) is that, at least asymptotically, the above anomalous dimensions must be negative. In the context of QCD similar negativity properties are a well-known fact, going by the name of the Nachtmann theorem \cite{Nachtmann:1973mr}.

Recently, a generalization of (\ref{eq:pos}) was proven in \cite{Kundu:2020gkz}, allowing one to make statements of OPEs involving different operators. If we denote by $\tau^*_{i j}(\ell)$ the minimal twist contained in the OPE of $\mathcal{O}_i$ with $\mathcal{O}_j$, the following inequalities have to be satisfied, for even spins $\ell_e \geq 2$ and odd spins $\ell_o \geq 3$ respectively:
\begin{equation}
\tau^*_{1 2}(\ell_e) \geq \frac{1}{2} \Big( \tau^*_{1 1}(\ell_e) +\tau^*_{2 2}(\ell_e) \Big)
\end{equation}
\begin{equation}
\tau^*_{1 2}(\ell_o) \geq \frac{1}{2} \Bigg( \frac{\tau^*_{1 1}(\ell_o+1) +\tau^*_{2 2}(\ell_o+1)}{2} +\frac{\tau^*_{1 1}(\ell_o-1) +\tau^*_{2 2}(\ell_o-1)}{2}  \Bigg)
\end{equation}
Although this seems to point in the direction of what we aim to establish, these bounds are automatically satisfied at leading order in $1/\ell$, for all choices of OPE coefficients and anomalous dimensions.

Finally, similar results were proven in a totally different context \cite{Hartman:2015lfa} using causality bulk techniques, by examining the two point function of an arbitrary operator $\mathcal{O}$ over a shock-wave background. There, the author argues that in the limit where the stress energy tensor is decoupled
\begin{equation}
\gamma(0,2) < 0 \quad \quad \text{for} \quad \quad \mathcal{O} \partial_{\mu} \partial_{\nu} \mathcal{O}.
\end{equation}
From a holographic point of view, this implies that a theory in AdS with the scalar Lagrangian
\begin{equation}\label{eq:phi4}
S = \int d^4x \sqrt{-g}  \big[ (\partial_{\mu} \varphi )^2 +\mu ( \partial_{\mu} \varphi)^4 \big]
\end{equation}
is only consistent if $\mathcal{\mu} > 0$.  This is the AdS generalization of a well-known result in flat spacetime \cite{Adams:2006sv} derived using the analyticity and causality structure of scattering amplitudes.

\subsubsection{Causality and analyticity  of scattering amplitudes}
As mentioned above, another form of sign constraint on EFTs comes from the study of consistency conditions on Effective Field Theories in Minkowski space arising from the combination of causality and S-matrix analiticity. Starting from \cite{Adams:2006sv}, it was shown that apparently healthy low energy theories cannot be UV completed into local quantum field theory or perturbative string theory if the signs of certain irrelevant operators are chosen appropriately. For simplicity, we only examine results for scalars, although there exist generalizations for non-trivial spins as well.

The simplest example is the theory of a shift symmetric, massless scalar $\pi$, which exemplifies the prototypical Nambu-Goldstone boson. The effective Lagrangian describing its self-interactions has the form
\begin{equation}
\mathcal{L}=(\partial \pi)^{2}+a \frac{(\partial \pi)^{2} \square \pi}{\Lambda^{3}}+c \frac{(\partial \pi)^{4}}{\Lambda^{4}}+\ldots
\end{equation}
where the coefficients of each term are naively unconstrained.\footnote{Although in specific cases their magnitude might be estimated, for example using Naive Dimensional Analysis (NDA).} Then, assuming the theory admits an expansion in a weak coupling $g$, it can be shown that the forward scattering amplitude $\mathcal{M}(s) = \mathcal{M}(s,t\rightarrow 0)$  admits a positive expansion, i.e. can be expanded in powers of $s$ as
\begin{equation}\label{eq:amp}
\mathcal{A}_{\text {tree }}(s)=g \sum_{n=1}^{\infty} c_{n}\left(\frac{s^{2}}{\Lambda^{4}}\right)^{n},
\end{equation}
in such a way that the coefficients always satisfy $c_n > 0$. This conclusion can be reached order by order by evaluating the complex contour integrals
\begin{equation}
I_{n}=\oint_{\gamma} \frac{d s}{2 \pi i} \frac{\mathcal{M}(s)}{s^{2 n+1}}
\end{equation}
along an infinite semi-circle lying on the axis, exploiting the analiticity of $\mathcal{M}(s)$. At this order in the coupling there can be no cuts in the amplitude, and all the poles give a positive contribution due to the Optical Theorem. To finish the argument, the Froissart bound $\mathcal{M}(s) < s \ln^2 s $ is used to kill the contribution at infinity for $n \geq 1$.

From an IR perspective, these bounds can also be understood to be closely connected with causality, since the wrong signs would yield faster than light propagations of fluctuations in the field over non trivial backgrounds.
This last aspect is best illustrated through a simple application: according to
(\ref{eq:amp}), a Lagrangian of the form
\begin{equation}
\mathcal{L}=  (\partial \pi)^{2}+ \mu \frac{(\partial \pi)^{4}}{\Lambda^{4}}
\end{equation}
requires $\mu > 0$, exactly as in Eq. (\ref{eq:phi4}).
If one quantizes over a non-zero background $ \partial_{\mu}\pi=C_{\mu} $, with $C^2 \ll \Lambda^4$, the linearised equations of motion imply the following dispersion relation for small excitations:
\begin{equation}
k^2- 4 \mu \frac{(C \cdot k)^2}{\Lambda^4} = 0,
\end{equation}
which clearly admits super-luminal modes for a negative $\mu$.

As a last remark, let us mention that there is one caveat to (\ref{eq:amp}), which amounts to the requirement that $\Lambda \ll M_P$. The reason is that if gravity is not weak enough (compared to the irrelevant operators) to be integrated out, any scattering amplitude will receive the universal contribution coming from tree level t-channel exchange
\begin{equation}
\mathcal{M}(s,t)\propto G_N \frac{s^2}{t}
\end{equation}
which diverges in the forward limit, invalidating the above discussion. From the point of view of causality, this happens because gravity bends all trajectories inside the lightcone \cite{Adams:2006sv}.

\section{String Compactifications and Holography}\label{sc:app}

\subsection{General considerations}
In this section we are concerned with relating swampland modifications on LVS and other low energy string compactification Lagrangians to constraints on QFTs and CFTs. A first crucial point to appreciate is that the established bounds do not directly lead to constraints for any of the stringy models under consideration.

From the perspective of the dual CFT, some difficulties arise when the bounds (\ref{eq:pos}) are applied in presence of gravity. One reason is that, as has already been mentioned, the actual minimal twist operators constrained by causality involve combinations of stress energy tensors, and not of the scalars. Secondly, the behaviour of the anomalous dimensions for large spin, i.e. where computations are most easily carried out, is dominated by graviton exchanges, saturating the $\ell$ exponent of the leading order term. If scalar couplings were all parametrically stronger than gravity, then it may be possible to neglect the gravitational exchanges. However, since moduli typically couple with gravitational-like strength it is not clear that it would be consistent to attempt to neglect dynamical gravity.

In the case of double trace operators built out of identical fields,\footnote{$[\mathcal{O}_{\varphi} \mathcal{O}_{\varphi}]_{n,\ell}$ and $[\mathcal{O}_{a} \mathcal{O}_{a}]_{n,\ell}$ in the case at hand.} negativity of the large-$\ell$ anomalous dimension is automatically satisfied and so does not lead to constraints: the sign is always determined by the product of the couplings, which are identical when the external legs are the same: hence $\gamma(n,\ell) \propto - g^2$. On the other hand, the constraints for different fields are always satisfied at large $\ell$.

A further limitation is that contact diagrams only give rise to tree-level anomalous dimensions up to a finite value of the spin, while exchanges contribute for any value of $\ell$. In particular, a 4-point interaction with $2j$ derivatives only contributes to anomalous dimensions $\gamma(n,\ell)$ with $n \leq j $ because of angular momentum conservation \cite{Heemskerk:2009pn,Fitzpatrick:2010zm}.
For exchange diagrams, the precise statement is that a 4-point Witten diagram, say in s-channel, contains an infinite number of spins when decomposed in conformal blocks in the t and u channels \cite{Zhou:2018sfz}.\footnote{Notice that here the same terminology (s,t and u) is used for two conceptually distinct concepts; the former usage refers to the form of the diagram, while the latter to the different OPE contractions used to expand the correlator.}
As the contact operators in the LVS Lagrangian have $\ell_c \le 1$ and bounds on anomalous dimensions are only valid for $\ell_c \geq 2$, positivity constraints do not apply directly to such operators.

It is interesting to notice that similar difficulties would appear if one tried to use the analyticity bounds to constrain the same coefficients for LVS in flat space-time, in the limit where $R_{AdS} $ is sent to infinity. In that case, only interactions with at least four derivatives are constrained since the bounds on the amplitude start at order $s^2$. Moreover, as observed in the previous section, the inclusion of dynamical gravity presents some obstructions to the use of these bounds.

\subsection{Holographic Analysis for LVS and Other Scenarios}

Having laid out these caveats, we now
want to determine some basic properties of LVS and other moduli stabilisation scenarios when viewed from a holographic perspective.

\subsubsection{Three point functions}\label{ssc:3ptf}

One fundamental computation, which will prove useful later, is that of the three point functions of the single trace scalars. As mentioned in section \ref{sc:boot}, these are fully constrained by conformal symmetry up to one coefficient $f_{ij}^k$.
To include all of our successive examples, we consider a generic theory containing a modulus and an axion, whose interactions can be parametrized up to cubic order as 
 \begin{equation}\label{eq:lagc}
\mathcal{L} \supset \frac{g}{M_P R^2_{AdS}}\varphi^3-  \frac{\mu}{M_P} \varphi\, \partial_{\mu} a \partial^{\mu} a
+ \frac{\kappa}{M_P R_{AdS}^2} \varphi a^2.
\end{equation}
The last coupling involving $\varphi a^2$ is absent in the basic LVS scenario, but is present in other scenarios (such as KKLT).
Because of the discrete symmetry $a \rightarrow - a$, the only non-trivial coefficients are $f_{\varphi \varphi}^{\varphi}$ and $f_{a a}^{\varphi}$ (parity implies $f^{\varphi}_{\varphi a} = f^a_{aa} = 0$).

We shall adopt the embedding formalism \cite{Penedones:2010ue,Kaplan}: points of $AdS_{d+1}$ are represented as null rays in $\mathbb{R}^{d+2,2}$,
\begin{equation}
X_0^2-X_1^2\,...\,-X_d^2+X^2_{d+1} = 1,
\end{equation}
on which the conformal group $SO(d,2)$ has a natural action.
The boundary is identified with the projective null cone
\begin{equation}
P^A \simeq \lambda P^A, \quad \quad \lambda \in \mathbb{R}.
\end{equation}
CFT operators can then be extended on the full space by imposing the homogeneity property
\begin{equation}
\mathcal{O}(\lambda P^A) = \lambda^{-\Delta} \mathcal{O}( P^A),
\end{equation}
and conformal invariance becomes manifest.  In this formalism, the bulk-to-boundary propagator of an operator with dimension $\Delta$ acquires the simple form
\begin{equation}
\Pi(P,X)_{\Delta} = \frac{1}{(2 P \cdot X)^{\Delta}}.
\end{equation}
The 3-point correlator arising from the $\varphi^3$ contact Witten diagram can then be written as
\begin{equation}\label{eq:3p}
\langle \varphi(x_1) \varphi(x_2) \varphi(x_3) \rangle = g \int_{AdS} \text{dX} \frac{1}{(2P_1 \cdot X)^{\Delta_{\varphi}} (2P_2 \cdot X)^{\Delta_{\varphi}}( 2P_3 \cdot X)^{\Delta_{\varphi}}}.
\end{equation}
Using the integral representation formula
\begin{equation}
\frac{1}{(2P \cdot X)^{\Delta}} = \frac{1}{\Gamma(\Delta)}\int \text{ds} \,  s^{\Delta} e^{-2 s P \cdot X},
\end{equation}
it follows (e.g. see \cite{Penedones:2016voo}) that the correlator has the form of Eq. (\ref{eq:3f}), with a coefficient
\begin{equation}
f_{\varphi \varphi}^{\varphi} = \frac{3g \pi^{d/2}  \Gamma \big( \frac{3 \Delta_{\varphi}-d}{2} \big)}{ \Gamma (\Delta_{\varphi})^3}.
\end{equation}
Note that, for $g>0$, this is positive for all $\Delta_{\varphi} > 1$.

For the correlation function involving axions, the appearance of derivatives can be translated into
\begin{equation}
\begin{split}
\langle \varphi(x_1) a(x_2)& a(x_3) \rangle = - \mu \int_{AdS} \text{dX} \frac{1}{(2P_1 \cdot X)^{\Delta_{\varphi}} } \\ & \times (\eta^{AB}+ X^A X^B) \frac{\partial}{\partial X^A} \frac{1}{(2P_2 \cdot X)^{\Delta_a}} \frac{\partial}{\partial X^B} \frac{1}{(2P_3 \cdot X)^{\Delta_a}}.
\end{split}
\end{equation}
With a few manipulations, and using $-P_1 \cdot P_2 = x_{12}^2$, the integral can now be turned into a sum of terms with the same structure as (\ref{eq:3p}), resulting in
\begin{equation}
f_{a a}^{\varphi} = \frac{ \pi^{d/2}  \Gamma \big( \frac{2 \Delta_a+ \Delta_{\varphi}-d}{2} \big)}{\Gamma (\Delta_{\varphi}) \Gamma (\Delta_{a})^2} \big[ \mu (\Delta_{\varphi}+ 2 \Delta_a -\Delta_a^2-3)+ \kappa \big].
\end{equation}
For future reference, we note that when $\Delta_a = 3$ and $\kappa = 0$ (corresponding to an axion with a flat potential, as in LVS) this becomes
\be
f_{a a}^{\varphi} = \frac{\mu \pi^{d/2}  \Gamma \big( \frac{2 \Delta_a+ \Delta_{\varphi}-d}{2} \big)}{\Gamma (\Delta_{\varphi}) \Gamma (\Delta_{a})^2} (\Delta_{\varphi} - 6),
\ee
a quantity which changes sign for $\Delta_{\varphi} > 6$.

\subsubsection{Mixed anomalous dimensions}\label{ssc:ma}

We reviewed in section \ref{sc:boot} how consistency constraints on EFTs often involve the signs of certain operators, and so we will be interested in quantities within the CFT that are sensitive to such signs.
One object in particular that we will be concerned with are the anomalous dimensions of mixed double-trace states, namely operators of the form  $[\mathcal{O}_{\varphi} \mathcal{O}_a]_{n,\ell}$. Although no \emph{a priori} constraints are known on their sign, we will see that for these operators negativity of the anomalous dimension appears to be crucial for satisfying many swampland conjectures.

From the point of view of the Analytic Bootstrap \cite{Komargodski:2012ek,Fitzpatrick:2012yx}, crossing symmetry does require in this context one specific behaviour for the anomalous dimensions as a function of the spin. Given two primaries $\mathcal{O}_1$ and $\mathcal{O}_2$ with twists $\tau_{\mathcal{O}_1}$ and $\tau_{\mathcal{O}_2}$, the bootstrap equations in Lorentzian signature\footnote{$d > 2$ is also assumed.} imply the existence of an infinite series of operators whose twist is arbitrarily close to the sum of the two. Furthermore, the twist of these operators asymptotically behaves as
\begin{equation}\label{eq:abts}
\tau(\ell) \simeq  \tau_{\mathcal{O}_1}+\tau_{\mathcal{O}_2} -\frac{c}{\ell^{\tau^*}} \textrm{\, \, for \, \,} \quad \quad \ell \gg 1,
 \end{equation}
where $\tau^*$ is the minimal twist amongst all operators contained in both the OPEs of $\mathcal{O}_1$ and $\mathcal{O}_2$ with themselves. While these results are easily seen to be a consequence of (\ref{eq:adf}) at tree level (see below), we emphasize that they apply in full generality.

\begin{figure}[h!]
\centering
\includegraphics[width=0.7\textwidth]{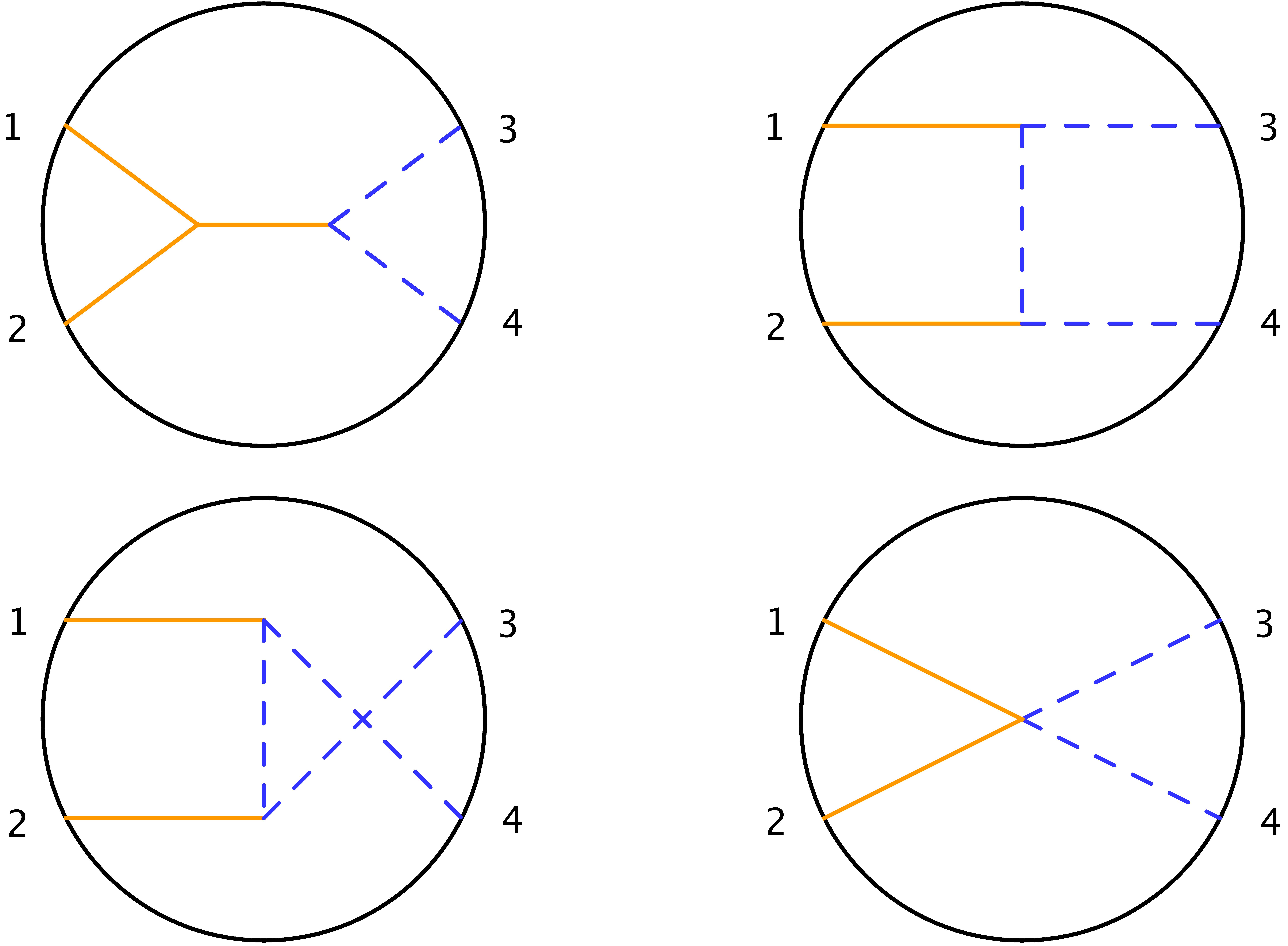}
\caption{Witten diagrams contributing to the anomalous dimensions of double trace operators of the form $[\mathcal{O}_{\varphi_1} \mathcal{O}_{\varphi_2}]_{(n,\ell)}$. The volume modulus corresponds to the continuous orange lines, and the axion to the blue dashed ones. The last diagram, without internal propagators, only contributes to anomalous dimension for small $\ell$.}
\label{fig:WittDiag}
\end{figure}

Assuming $\Delta_{\varphi} + \Delta_{a} \notin \mathcal{Z}$, the only tree-level Witten diagrams contributing to the mixed anomalous dimension are those associated with the correlation function
\begin{equation}
G(x_i) = \langle \mathcal{O}_{\varphi}(x_1) \mathcal{O}_{\varphi}(x_2) \mathcal{O}_a(x_3)  \mathcal{O}_a(x_4) \rangle,
\end{equation}
 and they are shown in Figure \ref{fig:WittDiag}. Furthermore, we make the claim that at $\mathcal{O}(\frac{1}{N^2})$, the leading contribution in the $1/\ell$ expansion only comes from the s-channel diagram.
Let us first give a generic argument based on the claim in the first paragraph, before reaching the same conclusion with the formula described in the last section. One can consider the following example: three scalars on AdS, with the only cubic vertex of the form
\begin{equation}
\mathcal{L} \supset g \varphi_1 \partial...\partial \varphi_2  \partial...\partial  \Phi,
\end{equation}
where any possible combination of derivatives is allowed.
Since vertices with more than three fields do not affect the anomalous dimension for large $\ell$ at this order in the coupling expansion, they can all be set to zero for the sake of the discussion. The Witten diagram $\varphi_1 \varphi_1 \rightarrow \varphi_2 \varphi_2 $ only contains t and u channels with an exchange of a $\Phi$, at order $1/N^2$.  Now, neither of the OPEs $\mc{O}_{\varphi_1} \mc{O}_{\varphi_1}$ or
$\mc{O}_{\varphi_2} \mc{O}_{\varphi_2}$
contains $\Phi$ at any order in perturbation theory, since the correlator $\langle  \mathcal{O}_{\varphi_1} \mathcal{O}_{\varphi_1} \mathcal{O}_{\Phi} \rangle$ vanishes as a result of the discrete symmetry
 \begin{equation}
 \mathcal{P}_{1}: \quad \varphi_1 \longrightarrow -\varphi_1, \quad \quad \quad \quad \Phi \longrightarrow -\Phi,
 \end{equation}
and the same result holds for $\mathcal{O}_{\varphi_2}$. More generally, the two OPEs do not contain \emph{any} common single trace operators.
It is thus possible to conclude that there is no contribution to the anomalous dimensions at leading order in the coupling from scalar diagrams built out of three-field interactions in $t$ and $u$ channel. The first correction will be due to the presence of the double trace operators $[\mathcal{O}_{\varphi_1} \mathcal{O}_{\varphi_1}]_{0,\ell}$\footnote{Assuming (without loss of generality) $\Delta_1 < \Delta_2 $.} in the OPE $\mathcal{O}_{\varphi_2} \times \mathcal{O}_{\varphi_2} $,
which results in a scaling behaviour of $1/\ell^{2 \Delta_1}$ at $\mathcal{O}\big( \frac{1}{N^4})$.
The intuitive meaning of this result is that the mixed anomalous dimension for the $\mc{O}_{\varphi_1} \mc{O}_{\varphi_2}$ operator is equivalent to the binding energy of the two particle $\varphi_1 \varphi_2$ state in AdS. This binding enegy would arise from particle exchange, which would correspond to s-channel for the $\varphi_1 \varphi_1 \to \varphi_2 \varphi_2$ topology.

A more direct way to see this is through the formula
\begin{equation}\label{eq:fa}
\begin{split}
\gamma(0,\ell) = - \int_{-i \infty}^{+ i \infty} & \frac{ds}{2 \pi i } M(s,0)\,\, _3F_2(-\ell, \Delta_1+\Delta_3+\ell -1,\frac{s}{2}; \Delta_1, \Delta_3;1 ) \\ & \times \Gamma \Big(\Delta_1-\frac{s}{2}\Big)\, \Gamma \Big(\Delta_3-\frac{s}{2} \Big)\, \Gamma \Big(\frac{s}{2}\Big)^2,
\end{split}
\end{equation}
where $M(s,t)$ is the Mellin amplitude defined with the conventions of paragraph \ref{eq:adf}.\footnote{There is a constant shift with respect to the canonical definition of the $t$ variable in order to simplify the form of the Mack polynomials. }.
For large $\ell$, it is possible to make the approximation
\begin{equation}
_3F_2(-\ell, \Delta_1+\Delta_3 +\ell-1,\frac{s}{2}; \Delta_1, \Delta_3;1 ) \simeq \frac{\Gamma(\Delta_1) \Gamma(\Delta_3)}{\Gamma(\Delta_1-\frac{s}{2})\Gamma(\Delta_3-\frac{s}{2})} \frac{1}{\ell^{s}}
\end{equation}
Given that the contour is closed on the right half-plane, the integral is dominated by the lowest pole in $s$ of $M(s,0)$ - that is the conformal dimension of the exchanged scalar. For a generic exchange diagram, the Mellin amplitude will have the form
\begin{equation}
M(s,t) = f_{12 X} f_{X34} \sum_m \frac{Q_{J_X,m}(t)}{s-\Delta_X+J_X-2m}+...
\end{equation}
where contributions from the crossed channels have been omitted.
In our case, the t-poles are set to zero and the only relevant ones are those coming from the  $s$ and $u$ channels. The latter, however, translate into a series of $s$ poles all on the left of the integration contour, because of the relation
\begin{equation}
s+t+u = \sum_i \Delta_i,
\end{equation}
and as such do not give any contribution to the integral. Therefore, the anomalous dimensions will only be sensitive to the s-channel amplitude, and scale as $1/\ell^{\Delta_{ex}}$ in the large $\ell$ limit - we therefore recover a special case of (\ref{eq:abts}). Moreover, equation (\ref{eq:fa}) easily allows to go one step further and actually compute the large-$\ell$ behaviour of $\gamma(0,\ell)$. Specialising to a Lagrangian of the form (\ref{eq:lagc}),
\begin{equation}
\label{xyz12}
\gamma^{\varphi a}(0,\ell) = -2 f_{\varphi \varphi \varphi} f_{\varphi a a }  \frac{  \Gamma(\Delta_a) \Gamma(\Delta_{\varphi})^2}{\Gamma \Big( \frac{2 \Delta_{a}-\Delta_{\varphi}}{2} \Big)\Gamma \Big( \frac{\Delta_{\varphi}}{2} \Big)^3}  \frac{1}{\ell^{\Delta_{\varphi}}}+ \mathcal{O}\Big(\frac{1}{\ell}\Big)
\end{equation}
which agrees with the formulas presented in \cite{Li:2015rfa}. Although their derivation is significantly shorter and does not require the introduction of Mellin amplitudes, we emphasize that our formalism provides a systematic procedure to calculate corrections at each order in $1/ \ell$. In principle, it is also possible to resum all these contributions and derive the value of the anomalous dimensions for finite spin, up to the effect of contact terms. The asymptotic series is guaranteed to converge for any spin $\ell>1$ thanks to the analiticity properties proven in \cite{Caron-Huot:2017vep}. Furthermore, we hope this approach might help to highlight possible connections with the standard bounds for scattering amplitudes in Minkowski space.

We flag up here the factor of
$$\frac{1}{\Gamma \Big( \frac{2 \Delta_{a}-\Delta_{\varphi}}{2} \Big)}$$
in Eq. (\ref{xyz12}), as for the case of light axions with $\Delta_a = 3$ this has interesting sign dependence on $\Delta_{\varphi}$.
\subsection{Implications}

We now study the anomalous dimensions within different scenarios of moduli stabilisation.
\subsubsection{LVS}
For the case of LVS, there are no $\varphi a a$ couplings in the potential. Then, inserting the value of the three-point function OPE coefficients from Eq. \ref{ssc:3ptf} immediately leads to
\begin{equation}
\gamma^{\varphi a}(0,\ell) = - \frac{g \mu (\Delta_{\varphi}-6)}{  M_P^2 R_{AdS}^2 } \frac{3 \pi^3}{\Gamma(\Delta_{\varphi})^2}\frac{\Gamma \big(\frac{3 \Delta_{\varphi}-3}{2} \big) \Gamma \big(\frac{\Delta_{\varphi}+3}{2} \big)}{  \Gamma \big(\frac{6 - \Delta_{\varphi}}{2} \big)
\Gamma^3 \big( \frac{\Delta_{\varphi}}{2}\big)} \frac{1}{\ell^{\Delta_{\varphi}} } + \mathcal{O}\Big(\frac{1}{\ell}\Big)
\end{equation}
where we have also specialized to $\Delta_a = 3$ and $d=3$.
As all other terms are automatically positive and $\Delta_{\varphi} \geq \frac{3}{2}$ due to the standard unitarity bound.\footnote{This is only true in absence of dynamical gravity, which induces correction saturating the unitarity bound on the $1/\ell$ exponent and thus scaling as $1/ \ell^{d-2}$. Since the moduli interactions are suppressed by the same powers of the Planck mass as standard gravitational interactions, it is only possible to integrate out such effects if $g \mu >> 1$, and otherwise Eq. (\ref{eq:andim}) actually parametrizes to the leading behaviour of
\begin{equation}
\gamma^{\varphi a}(0,\ell)- \gamma^{\varphi a}(0,\ell)_{grav}
\end{equation}
for sufficiently high values of $\ell$.
In principle, one could also suspect that subleading terms in the $1/N$ expansion might decrease with a lower power of $\ell$ and dominate asymptotically if the coupling is fixed. In our example, however, $R_{AdS}$ diverges in the large volume limit and it is legitimate to consider the $\frac{1}{N^2}$ correction only.}, the sign of the $\phi a $ anomalous dimensions is governed, at least asymptotically, by the sign of the combination
\begin{equation}\label{eq:andim}
\gamma^{\varphi a}(0,\ell) \sim  -g \mu  \frac{\left( \Delta_{\varphi}-6 \right)}{\Gamma \big(\frac{6 - \Delta_{\varphi}}{2} \big) }.
\end{equation}

We can now consider the meaning of equation (\ref{eq:andim}) in more detail. In particular, an interesting correlation appears between the allowed values of the parameter space in String Theory and the sign of the anomalous dimension, which is highly reminiscent of the bounds discussed in the previous sections.

In particular, the RHS is negative if the variables take values in the original LVS model in the $\mc{V} \to \infty$ limit (with $\Delta_{\varphi} = 8.04$), but switches to positive if certain swampland-like transformation are performed on the AdS theory. The most obvious example of these is the sign exchange $\mu \rightarrow - \mu$,
corresponding on the string theory side (as discussed in \ref{ssc:swc}) to an axion decay constant that diverges in the large volume limit.
A similar behaviour occurs for the change $g \rightarrow - g$. A simultaneous change of both signs is allowed, as this is equivalent to the field redefinition $\varphi\rightarrow - \varphi$, but the product of the two signs is a physical quantity independent of the field redefinition.

\begin{figure}[h!]
\centering
\includegraphics[width=0.8\textwidth]{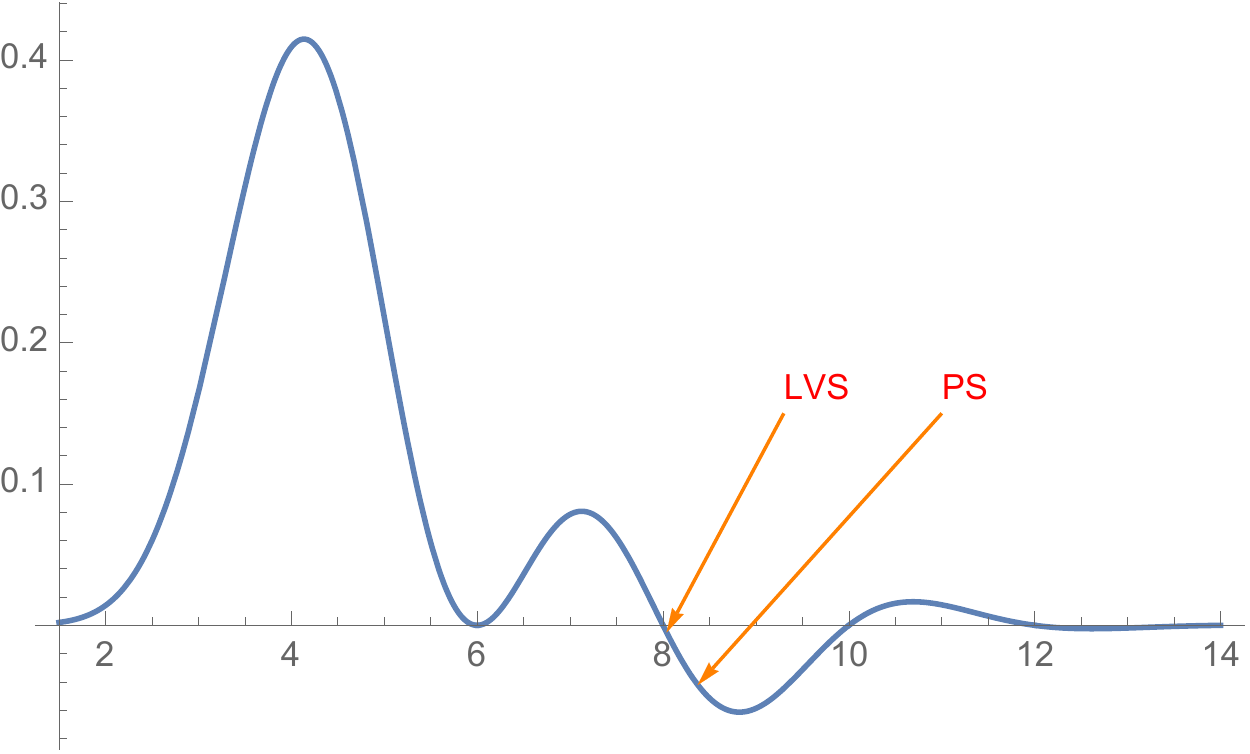}
\caption{Prefactor $C(\Delta_{\varphi})= \gamma(0,\ell) \ell^{\Delta_{\varphi}}$ as a function of $\Delta$. For graphical purposes, it has been rescaled by a factor of $\big(\frac{\Delta}{10}\big)^{6}$. The lowest possible value admitting a negative anomalous dimension is $\Delta =8$. From then onwards, the sign oscillates between positive and negative values with decreasing amplitude and period of 2. The two arrows show the location of the LVS and Perturbative Stabilization models respectively, in the $\mathcal{V} \rightarrow \infty$ limit .}
\label{fig:sign}
\end{figure}

There is also an interesting dependence on $\Delta_{\varphi}$, as
any value $\Delta_{\varphi} < 8$ (and consistent with the unitarity bound) also gives rise to a positive anomalous dimension, as can be seen in Fig.\ref{fig:sign}.
Recall that the value of $\Delta_{\varphi}$ arises from the volume scaling of the potential.
Quite remarkably, LVS is found to reside in a rather special place, right at the edge of the lower boundary of the `allowed' region with negative anomalous dimension.

While the region where $\Delta_{\varphi} < 6$ can be understood as corresponding to a potential growing parametrically faster than the string scale $M_s^4$, we have no valid interpretation as to why $\Delta_{\varphi} = 8$ should be regarded as a critical value. We also note that $\gamma(0,\ell)$ goes to zero exactly at the point where $\Delta_{\varphi} = 6$, which by the above argument is expected to exhibit some kind of transition.

It is intriguing that in the $\mathcal{V} \to \infty$ limit of LVS, the value of $\Delta_{\varphi}$ is so close to 8. At finite volume,
$\Delta_{\varphi} $ approaches the asymptotic value of $\frac{3}{2}\big(1+\sqrt{19}\big) \simeq 8.038$ from below as the volume is brought into infinity, with corrections that scale as $\delta \Delta_{\varphi} \sim \frac{1}{\ln \mc{V}}$.
Including these corrections in the 1-modulus potential explicitly, one finds that the finite volume necessary to achieve $\Delta_{\varphi}=8$ is given by
\begin{equation}
\langle \varphi \rangle = \frac{1}{2 \big(\lambda-\frac{40}{3 \lambda} \big)}
\end{equation}
Numerically, this amounts to $\mathcal{V} \sim 10^5/10^6$.

\begin{figure}[h!]
	\label{figxyz}
	\centering
	\includegraphics[width=0.45\textwidth]{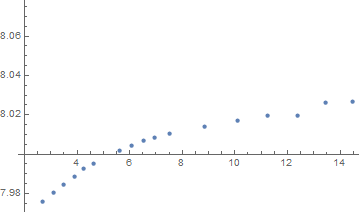} \includegraphics[width=0.45\textwidth]{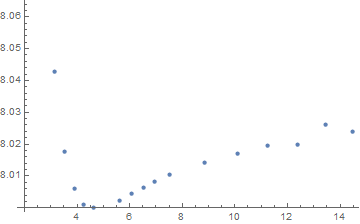}
	\caption{ This plot illustratcaes the effect of a higher-order correction to the K\"ahler potential. The x-axis shows the logarithm (in base 10) of the stabilised volume while the y-axis shows the conformal dimension of the dual volume operator, $\Delta_{\varphi}$. The case on the left is LVS based on the original $\mathbb{P}^4_{1,1,1,6,9}$, with a volume $\mc{V} = \frac{1}{9\sqrt{2}}\left( \tau_b^{3/2} - \tau_s^{3/2} \right)$, a K\"ahler potential $- 2 \ln \left( \mc{V} + \xi \right)$, and a superpotential $W = W_0 + A_s e^{-2 \pi T_S}$ with $W_0 = 1$ and $A_s = 1$. Different values of the stabilised volume have been obtained by taking different values of $\xi$. In this case, the logarithmic corrections are seen to bring $\Delta_{\varphi}$ below 8 at smaller volumes. The case on the right is identical, except for a K\"ahler potential
		$K = - 2 \ln \left( \mc{V} + \xi + \frac{\xi^2}{\mc{V}} \right).$ In this case, $\Delta_{\varphi} > 8$ throughout.}
	\label{fig:plot1}
\end{figure}

However, one cannot conclude that at such volumes it is definitively true that $\Delta_{\varphi} < 8$, as once volumes become smaller
corrections that are higher-order in $\alpha'$ (and so can normally be neglected) become relevant for the question of whether $\Delta_{\varphi} > 8$ or $\Delta_{\varphi} < 8$. Specifically, the fractional difference between 7.99 and 8.00 is $\mc{O}(10^{-3})$ and so effects that are far too small to modify the existence of the minimum are capable of changing the sign of $(\Delta_{\varphi} - 8)$.
This can be verified by a numerical study of the full 2-modulus LVS stabilisation. Here one sees that the sign of $\Delta_{\varphi} - 8$ at finite $\mc{V}$ (e.g. $\mc{V} \sim 10^6$) can easily be modified by corrections of the form
$$
K = - 2 \ln \left( \mc{V} + \xi \right) \to K = - 2 \ln \left( \mc{V} + \xi + \frac{\xi^2}{\mc{V}} \right).
$$
This is illustrated in figure \ref{fig:plot1}.
As the determination of the coefficients of such corrections is beyond any current control, this demonstrates why there is no controlled region of LVS in which one can be sure that $\Delta_{\varphi} < 8$.

It is suggestive to find a connection between the Swampland and something so closely resembling CFT consistency conditions. Furthermore, the $\gamma^{\varphi a}$'s are essentially the only OPE data (at least at tree-level) sensitive to the sign of the product $g \mu$. However we stress again that negativity of the anomalous dimensions is not established in general for double trace operators constructed of two different primaries.

\subsubsection{Perturbative Stabilisation}

We can perform a similar analysis for other models of IIB moduli stabilisation, even if they are not be as fully worked out as LVS.
In particular, they do not offer the cleanliness of the $\mc{V} \to \infty$ limit of LVS (the holographic study of IIA moduli stabilisation scenarios \cite{DeWolfe:2005uu} would also be interesting, as it offers a different large-volume limit driven by large flux quanta, but beyond the scope of this work).
One example is that of the so-called perturbative stabilization \cite{Berg:2005yu,vonGersdorff:2005bf}, in which the volume is stabilised by the competing effects of two separate perturbative corrections to the scalar potential. Here the dynamics of the volume modulus is governed by a potential of the form
\begin{equation}
V(\varphi)= A e^{-\frac{\lambda_1 \varphi}{M_P}}- B e^{-\frac{\lambda_2 \varphi}{M_P}},
\end{equation}
which reflects competition between two effects scaling with different powers of the volume.
Such a potential could, for instance, be generated by a combination of $\alpha'$ corrections scaling as $\mathcal{V}^{-3}$, like in LVS, and string loop effects of order $\mathcal{V}^{-\frac{10}{3}}$. The minimum is located at
\begin{equation}
\varphi_c = \frac{\log \Big( \frac{A \lambda_1}{B \lambda_2} \Big)}{\lambda_1 - \lambda_2}, \quad \quad \text{with } \quad \quad V_{min}= B e^{-\lambda_2 \varphi_c} \left( \frac{\lambda_2}{\lambda_1}-1 \right),
\end{equation}
and results in an AdS vacuum for $\lambda_1 > \lambda_2$. In a top-down scenario, the coefficients $\lambda_1$ and $\lambda_2$ would be fixed. However, for purpose of exploration we may consider them arbitrary, allowing for the case that $\vert \lambda_1 - \lambda_2 \vert \ll \vert \lambda_1 + \lambda_2 \vert$ which leads to large volumes (alternatively, one could demand that the coefficients are fine tuned with $A \gg B$).
Expanding about the minimum, one recovers n-point interactions for the volume modulus
\begin{equation}
\mathcal{L}_{ {\varphi}^n}= (-1)^{n-1}\frac{3M_P^2 \lambda_1 \lambda_2}{R_{AdS}^2} \frac{\lambda_1^{n-1}-\lambda_2^{n-1}}{\lambda_1 - \lambda_2}
( {\varphi})^n.
\end{equation}
Consequently the conformal dimension of the corresponding CFT operator is
\begin{equation}
\Delta_{\varphi}= \frac{3(1 \pm \sqrt{1+\frac{4}{3}\lambda_1 \lambda_2})}{2}.
\end{equation}
The resemblance with the result for LVS should not be too surprising, as the potential actually reduces to that of standard LVS in the limit $\lambda_1 \rightarrow \lambda_2$. The generalization of Eq. (\ref{eq:andim}) is
\begin{equation}
\gamma^{\varphi a}(0,\ell) \sim -\mu \lambda_1 \lambda_2(\lambda_1+\lambda_2)\frac{\left( \Delta_{\varphi}-6 \right)}{\Gamma \big(\frac{6 - \Delta_{\varphi}}{2} \big) } ,
\end{equation}
which again lends itself to an interpretation in terms of Swampland constraints. For the values considered above, namely
\begin{equation}
\lambda_1 = \frac{10}{3}\sqrt{ \frac{3}{2}}, \quad \quad \quad
\lambda_2 = 3 \sqrt{ \frac{3}{2}},
\end{equation}
the anomalous dimensions are again negative. However, the sign changes under any of the following:
\begin{itemize}
\item $\mu \rightarrow - \mu$, corresponding to $f_a \sim M_P \mathcal{V}^{\frac{2}{3}}$.
\item $\lambda_2 \rightarrow - \lambda_2$, which leads to an unbounded potential for large values of the volume. Notice that the same does not happen if one inverts the sign of $\lambda_1$, since $\lambda_1 > \lambda_2$ and the anomalous dimension is proportional to the product $ (\lambda_1 + \lambda_2)$
\item $0 \leq \lambda_1 \lambda_2 \leq \frac{40}{3} $:
This is exactly analogous to LVS.
If $0 \leq \lambda_1 \lambda_2 \leq 6 $, there will be at least one term in the potential growing faster than $M_s^4$, but otherwise there is no clear interpretation. As a further comment, this is slightly reminiscent (but qualitatively different) from similar constraints arising from the De Sitter Swampland conjecture, of the schematic form $\lambda_i < c $ with $c$ an order one constant \cite{Obied:2018sgi}.
\end{itemize}

\subsubsection{KKLT}

Another well-studied case where these ideas can be tested is the KKLT construction \cite{Kachru:2003aw}.
Although this ultimately aims at de Sitter space, KKLT
starts with an AdS vacuum that is subsequently uplifted to dS: it is the former that we will discuss here.

The Kähler potential is given by the standard tree-level expression
\begin{equation}
K= -3 \ln [-i(\rho-\bar{\rho})],
\end{equation}
while the superpotential includes a tree level contribution from fluxes and  non-perturbative effects.
\begin{equation}
W= W_0+Ae^{i \alpha \rho}.
\end{equation}
The supergravity potential
\begin{equation}\label{eq:sugra}
V= e^{K} \big(G^{\rho \bar{\rho}} D_{\rho} W \bar{D_{\rho}} W - 3 \abs{W}^2 \big)
\end{equation}
can be minimised while simultaneously preserving supersymmetry, for a critical value of the modulus field $\sigma_{cr}$ satisfying
\begin{equation}
\frac{W_0}{A} = -e^{-\alpha \sigma_{cr}}(1+\frac{2 \alpha \sigma_{cr}}{3}).
\end{equation}
Since the supergravity approximation and the single instanton approximation
require $\sigma_{cr} \gg 1$  and $ \alpha \sigma_{cr} > 1$ respectively,  $W_0$ has to be tuned to be very small.

The situation when considering low energy excitations around this vacuum is qualitatively different from LVS.
As the instanton breaks the continuous shift symmetry, the axion now has both a mass and also new potential-type interactions with the volume modulus.
From a holographic perspective, the first consequence is that the conformal dimension of the axion is no longer $\Delta = 3$, but instead
\begin{equation}
\Delta_{a} = \frac{3}{2} \Bigg( 1 \pm \sqrt{1+\frac{8}{9}\alpha \sigma_{cr}(2 \alpha \sigma_{cr}+3 ) } \, \Bigg),
\end{equation}
while for the volume modulus, the result is
\begin{equation}
\Delta_{\varphi} = \frac{3}{2} \Bigg( 1 \pm \sqrt{1+\frac{8}{9}(2+\alpha \sigma_{cr})
(1+2 \alpha \sigma_{cr} )} \,  \Bigg).
\end{equation}
A noteworthy point is that the last two equations imply $\Delta \varphi < 2 \Delta a$, so that the oscillating sign behaviour due to the negative gamma function is not reproduced in this case.

The cubic interaction terms in the potential now read
\begin{equation}\label{eq:pkklt}
\begin{aligned}
V^{(3)}(\varphi,a) \supset & -\frac{1}{ M_P R^2_{AdS}}\sqrt{\frac{2}{3}} \big[ 2+ \alpha \sigma_{cr}(5+\alpha \sigma_{cr}(5+2\alpha \sigma_{cr}))\big] \varphi^3 \\
& -\frac{1}{ M_P R^2_{AdS} } \sqrt{\frac{2}{3}} \alpha \sigma_{cr}(2+\alpha \sigma_{cr})(3+2\alpha \sigma_{cr}) \varphi a^2.
\end{aligned}
\end{equation}
The interactions of the form $\varphi^n \partial_{\mu}a \partial^{\mu}a$ are left unchanged as they arise from the axion kinetic term.
For completeness, we also report the quartic interactions in the potential:
\begin{equation}
\begin{aligned}
V^{(4)}(\varphi,a) \supset & \frac{1}{ 18M^2_P R^2_{AdS}} \Big[ (1+\alpha \sigma_{cr})(14+\alpha \sigma_{cr}(21+\alpha \sigma_{cr}(11+14\alpha \sigma_{cr})))\varphi^4
\\& + 6 \alpha \sigma_{cr}(3+2\alpha \sigma_{cr})(4+\alpha \sigma_{cr}(3+\alpha \sigma_{cr}))a^2 \varphi^2- 6 \alpha \sigma_{cr}^3(3+2 \alpha \sigma_{cr})a^4\Big].
\end{aligned}
\end{equation}

In particular, the anomalous dimensions can be computed as in section \ref{ssc:ma}
(note that the fact that $\Delta \varphi < 2 \Delta a$ also implies that the dominating contribution to anomalous dimensions at large $\ell$ will always be the one coming from the exchange of a $\varphi$, and one does not have to worry about potential loop effects proportional to $\ell^{-2 \Delta_a}$).
The main difference is that there are now two contributions to $f_{aa}^{\varphi}$, arising from both the $\varphi a^2 $ and $\varphi \partial_{\mu}a \partial^{\mu}a $ vertices, and with opposing signs. As the
sign of the $\varphi^3$ coefficient in (\ref{eq:pkklt}) is fixed
it is the sum of these two compensating effects that must be considered, according to
\begin{equation}
\gamma(0,\ell) \propto - g \Big[
 \alpha \sigma_{cr} (2+\alpha \sigma_{cr}) (3 + 2 \alpha \sigma_{cr}) +
  2 \big( \Delta_{\varphi}(\alpha \sigma_{cr})- \Delta_{a}^2(\alpha \sigma_{cr}) + 2\Delta_{a}(\alpha \sigma_{cr}) - 3\big) \Big].
\end{equation}
The anomalous dimensions are then negative provided
\begin{equation}
\alpha \sigma_{cr} \gtrsim 2.24,
\end{equation}
which is always true in KKLT's regime of validity of $W_0 \sim e^{-\alpha \sigma} \ll 1$ where any multi-instanton effects can be neglected.

We note that negativity was not automatic and
required the presence of the additional $\varphi a^2 $ term in the potential, as the values of the conformal dimensions were such that the contribution from the derivative vertex switched sign with respect to LVS and the potential contribution was necessary to produce negative anomalous dimensions.

\subsubsection{Racetrack}

Another popular scenario has been that of racetrack stabilisation, where the tree-level flux superpotential is set to zero and the dominant effect instead results from an interplay between two different non-perturbative effects (for example see \cite{Escoda:2003fa}),
\begin{equation}
W= A e^{i \alpha \rho}-B e^{i \beta \rho}.
\end{equation}
Assuming without loss of generality $\alpha > \beta > 0$ a supersymmetric minimum of the potential is found at
\begin{equation}
e^{-(\alpha-\beta)\sigma_{cr}} = \frac{B(3 + 2 \beta
 \sigma_{cr})}{A(3 + 2 \alpha \sigma_{cr})}.
\end{equation}
Using the same techniques, it is possible to compute the supergravity potential in Eq. (\ref{eq:sugra}) and expand around its minimum to obtain the conformal dimensions and low-point interactions of the axion and volume modulus.
\begin{equation}
\Delta_{a} = \frac{3}{2} \Bigg( 1 \pm \sqrt{1+\frac{8}{243} (3 + 2 \alpha \sigma_{cr}) (3 + 2 \beta
 \sigma_{cr}) (2 \alpha \beta \sigma_{cr}^2  + 3 \alpha \sigma_{cr} + 3 \beta
 \sigma_{cr}) } \,  \Bigg),
\end{equation}
\begin{equation}
\Delta_{\varphi} = \frac{3}{2} \Bigg( 1 \pm \sqrt{1+\frac{8}{243}
 (6 + 3 \alpha \sigma_{cr} + 3  \beta
 \sigma_{cr} + 2 \alpha \beta \sigma_{cr}^2 ) (3 + 4 \alpha \beta \sigma_{cr}^2 + 6 \alpha \sigma_{cr} +  6 \alpha\sigma_{cr})) } \, \Bigg).
\end{equation}
They can be shown to satisfy $\Delta_a < \Delta_{\varphi} < 2 \Delta_a$, so that again there is no oscillating behaviour and one should not worry about higher order effects becoming relevant for large $\ell$.

Here, the cubic terms in the potential read
\begin{equation}
\begin{aligned}
& V^{(3)}(\varphi,a) \supset  -\frac{\sqrt{\frac{2}{3}} }{ 9M_P R^2_{AdS}}  \Big[(\alpha  \sigma  (2 \beta  \sigma +3)+3 \beta  \sigma +3) (\sigma  (\alpha +\beta ) (2 \alpha  \sigma +3) (2 \beta  \sigma +3)+6)  \varphi^3 \\ 
&  -  \sigma  (2 \alpha  \sigma +3) (2 \beta  \sigma +3) \left(\alpha ^2 \sigma  (2 \beta  \sigma +3)+2 \alpha  (\beta  \sigma +1) (\beta  \sigma +3)+3 \beta  (\beta  \sigma +2)\right) \varphi a^2 \Big],
\end{aligned}
\end{equation}
while the quartic ones are
\begin{equation}
\begin{aligned}
V^{(4)}(\varphi,a) \supset & \frac{4}{ 162 M^2_P R^2_{AdS}} \Big[ \varphi^4  \sigma ^3 (\alpha -\beta )^2 (2 \alpha  \sigma +3) (2 \beta  \sigma +3) (2 \alpha  \beta  \sigma +3 (\alpha +\beta ))
\\& - 6 \sigma  (2 \alpha  \sigma +3) (2 \beta  \sigma +3) (\alpha ^3 \sigma ^2 (2 \beta  \sigma +3)+\alpha ^2 \sigma  (\beta  \sigma  (4 \beta  \sigma +11)+9) \\&+ \alpha  (\beta  \sigma +2)^2 (2 \beta  \sigma +3)+3 \beta  (\beta  \sigma  (\beta  \sigma +3)+4))a^2 \varphi^2 \\&
-\big( 14 \alpha ^4 \sigma ^4 (2 \beta  \sigma +3)^2 +5 \alpha ^3 \sigma ^3 (2 \beta  \sigma +3) (8 \beta  \sigma  (\beta  \sigma +3)+15)
\\&+\alpha ^2 \sigma ^2 (2 \beta  \sigma +3) (\beta  \sigma  (2 \beta  \sigma  (14 \beta  \sigma +69)+169)+96)\\&+3 \alpha  \sigma  (2 \beta  \sigma +3) (\beta  \sigma  (\beta  \sigma  (28 \beta  \sigma +43)+52)+35)\\& +9 (\beta  \sigma +1) (\beta  \sigma  (\beta  \sigma  (14 \beta  \sigma +11)+21)+14) \big) a^4\Big].
\end{aligned}
\end{equation}
This time, the sign of the anomalous dimensions is determined by the combination
\begin{equation}
\begin{aligned}
\gamma(0,\ell)&  \propto  -{\mu} \, \sqrt{\frac{2}{
  3}} \,  \Big[ \frac{1}{9} (3 + 2 \beta \sigma_{cr}) (3 + 2  \alpha \sigma_{cr}) (
  2\alpha^2\beta \sigma_{cr}^3 + 2\alpha \beta^2\sigma_{cr}^3 \\ &
  +3 \alpha^2 \sigma_{cr}^2 + 3 \beta^2 \sigma_{cr}^2+ 8 \alpha \beta  \sigma_{cr}^2  + 2\alpha \beta^2\sigma_{cr}^2 + 6 \alpha  \sigma_{cr} + 6 \beta  \sigma_{cr} )\\ & +
  2  \big( \Delta_{\sigma}(\alpha \sigma_{cr},\beta \sigma_{cr})- \Delta_a^2( \alpha \sigma_{cr},\beta \sigma_{cr}) +
    2 \Delta_a(\alpha \sigma_{cr}, \beta \sigma_{cr}) - 3\big) \Big].
\end{aligned}
\end{equation}
Numerically, one can verify that this is negative in the controlled region $(\alpha+\beta) \sigma_{cr} \gtrsim 0.75 $, which again contains the region of parameter space where the single-instanton approximation is valid for both terms in the superpotential.

The summary of this is that in a variety of examples negativity of $\gamma_{\varphi a}$ is satisfied within the controlled region, with a variety of interlocking parts leading to this conclusion.

\section{Heavy States and the Swampland Distance Conjecture}
\label{sec5}

So far our swampland analysis has focused on the question of transPlanckian axion decay constants. However we can also
make connections to the swampland distance conjectures \cite{Ooguri:2006in,Klaewer:2016kiy}. These conjectures are statements about the appearance of
light towers of states as one moves through large (in general transPlanckian) distances in moduli space. They come
in both original \cite{Ooguri:2006in} and refined versions \cite{Klaewer:2016kiy}.

The original version states that as one moves an asymptotic geodesic distances $d(P, Q)$ (from $P$ to $Q$ in moduli space), a tower of states becomes light, with the masses of the tower particles descending as
$$
M(P) = M(Q) e^{-\lambda d(P, \, Q)/ M_P}
$$
in the limit that $d(P, Q) \to \infty$. Examples of this behaviour are the towers of Kaluza-Klein or string modes that become light on
 moving to asymptotically large geometric volumes, or alternatively the towers of winding/wrapped brane modes in the limit that cycle sizes collapse to zero volume.

The refined version states that the constant $\lambda$ is $\mathcal{O}(1)$ and that the behaviour is realised not merely asymptotically but already once $d(P, Q) \gtrsim M_P$. Equivalently, the tower of light modes occurs not simply at the asymptotic boundaries of moduli space but already once transPlanckian field displacements occur.

For the purpose of constraining low-energy effective field theory Lagrangians, it is the refined swampland distance conjecture that is much more powerful. The original swampland distance conjecture constrains behaviour at asymptotic distances in field space. However, the low-energy physicist is only sensitive to a particular EFT and perturbations around it, while such asymptotic displacements are not accessible as a small perturbation about any low-energy Lagrangian -- and so cannot be used to say that a particular effective field theory is or is not in the swampland.

This is not so for the refined distance conjecture. The couplings and interactions of gravitationally coupled
moduli $\varphi$ can be ordered by an expansion in $\left(\frac{\varphi}{M_P} \right)$, $\left( \frac{\varphi}{M_P} \right)^2$.
Constraints on the behaviour of masses (for example) for transPlanckian moduli displacement constrain the form and signs of such couplings,
as it is essential to the refined distance conjecture that the behaviour kick in once $\Delta \varphi \sim M_P$.
This therefore provides constraints both on models of large-field inflation and also on the leading perturbative interactions of gravitationally coupled scalars (such as moduli).

\subsection{A Puzzle}\label{ssc:pzl}

Although a lot of evidence exists for some form of the refined distance conjecture \cite{Blumenhagen:2017cxt,Blumenhagen:2018nts,Erkinger:2019umg}, at this point we want to
note a puzzle concerning the refined distance conjecture. The conjecture states that towers of states 
should become light for geodesic displacements $\Delta \varphi \gtrsim M_P$ in moduli space. We consider an ordinary  type IIB Calabi-Yau orientifold compactification in a limit of large internal geometric internal volume $\mc{V} \gg l_s^6$ (such as LVS). As the canonically normalised volume field is $\varphi = M_P \sqrt{\frac{3}{2}} \ln \tau_b = M_P \sqrt{\frac{2}{3}} \ln \left( \mc{V}/l_s^6 \right)$, large field displacements ($\vert \Delta \varphi \vert \gg M_P$) can be achieved by exponential rescalings $\mc{V} \to e^{\pm \sqrt{\frac{3}{2}} \vert \Delta \varphi \vert / M_P} \mc{V}$.

For displacements with positive $\Delta \varphi > 0$, it is clear how the conjecture is satisfied.
In this case there are multiple towers of modes becoming light, for example the towers of string or KK modes.
Taking these as canonical examples, their masses behave as
$$
M_{string} \propto \frac{M_P}{\sqrt{\mc{V}}}, \qquad M_{KK} \propto \frac{M_P}{\mathcal{V}^{2/3}},
$$
using the standard relationship between the string scale and 4-d Planck scale. As $M < M_P$, these indeed correspond to particle states in 4d QFT.
For positive displacements $\Delta \varphi > 0$
the masses of the states in these towers then behave as
$$
M_{string} = e^{-\half \sqrt{\frac{3}{2}} \Delta \varphi / M_P}, \qquad  M_{KK} = e^{- \frac{2}{3} \sqrt{\frac{3}{2}} \Delta \varphi / M_P}.
$$

However, the refined distance conjecture refers simply to transPlanckian displacements $d(P,Q) > M_P$ and so
should therefore apply equally to displacements with $\Delta \varphi < 0$ -- corresponding to rescalings $\mc{V} \to e^{-\sqrt{\frac{3}{2}} \vert \Delta \varphi \vert} \mc{V}$. Equivalently, when writing an effective Lagrangian that is valid about
a large-volume locus in moduli space, we are free to make a field redefinition $\varphi \to -\varphi$, while any general statement about low energy effective Lagrangians must remain valid.

This is where the puzzle lies -- for if we start at large volumes $\mc{V} \ggg 1$ and move inwards in moduli space towards smaller volumes, with $\vert \Delta \varphi \vert > M_P$ but $\Delta \varphi < 0$, there is no apparent tower of particle states that becomes light.
It is clear that as $\mc{V}$ decreases the string and KK tower becomes heavier. What about the tower of winding states?
As these have masses $M_{winding} \sim R M_{string}$, it is true that $\frac{M_{winding}}{M_{string}}$ decreases as the volume decreases. However, as $M_s \sim \frac{M_P}{\sqrt{\mc{V}}}$, we have overall
$
M_{winding} \sim \frac{M_P}{\mc{V}^{1/3}},
$
and so the tower of winding states also \emph{increases} in mass under a displacement $\Delta \varphi < 0, \vert \Delta \varphi \vert > M_P$.

We can also consider states arising from wrapped branes. A D$(p+1)$-brane wrapped on an internal $p$-cycle $\Sigma_p$ corresponds to a particle state in spacetime with a mass $M \sim \left( {\textrm{Vol}(\Sigma_p)}/{\sqrt{\mc{V}}}\right) M_P$. For branes wrapping internal 1-cycles or 2-cycles, the mass of the resulting state increases under reductions in the bulk volume (as $\textrm{Vol}(\Sigma_p) \propto \mc{V}^{1/6}$  or  $\mc{V}^{1/3}$ for bulk 1- or 2-cycles). For a brane wrapping a internal 3-cycle, the volume factors cancel and so the tower remains unaltered in mass.

These volume scalings imply that a tower of particle states decreasing in mass under reductions in volume, if coming from wrapped branes,
would require the branes to wrap bulk 4-cycles (or 5-cycle or 6-cycles) in the internal space.
However, in the limit of large radii $R \gg l_s$ this is immediately problematic. For a brane wrapped on a 4-cycle,
the mass of such `states' behaves as $M \sim R M_P$, and so in a large radius limit such wrapped branes have masses above the 4-d Planck scale. As such, they cannot be interpreted as particles within the 4d effective field theory and instead correspond to black holes. In the large radius limit of $R \gg l_s$ they do not provide examples of particle states that become lighter on reduction of the internal volume.

It is still true that in the strict limit of $d(P, Q) \to \infty$ (as in the original distance conjecture), this is not an issue -- in the limit of infinite $d(P, Q)$ the compactification reaches down to (formally) zero radius, entering a regime where
winding modes become light and wrapped brane states do correspond to particles. However, this puzzle can be formulated
for any arbitrarily large but finite value of $d(P,Q)$: by working with exponentially large internal volumes, and starting arbitrarily far away from the centre of moduli space, the geodesic distance to the self-dual radius can be made arbitrarily large.

The large volume limit therefore appears to create a problem for the ordinary formulation of the refined swampland distance conjecture: under displacements $\vert \Delta \varphi \vert \gg M_P$, $\Delta \varphi < 0$, corresponding to exponential reductions in the internal volume, all the towers of heavy particle states appear to be \emph{increasing} in mass, with
$M_{tower} \to e^{+\lambda \vert \Delta_{\varphi} /M_P \vert} M_{tower}$, rather than decreasing.

\subsection{Heavy Modes and Holographic Anomalous Dimensions}

We now extend our earlier results involving the $[\varphi a]$ mixed correlator to the case of mixed correlators involving heavy modes. In particular, we examine the implications of requiring
a negative anomalous dimension for double trace CFT states in which one operator corresponds to a heavy field $\psi$. `Heavy' here is defined by a condition that the conformal dimension $\Delta_{\psi}$ of the field diverges in the limit that $\mc{V} \to \infty$. For the case of LVS, such modes can correspond either to certain moduli (such as the complex structure moduli or the small K\"ahler moduli) or to the KK, string or winding modes.

We restrict our analysis in this section to LVS.
There are then three types of mixed state involving a heavy mode that we can consider. These are mixed states containing
two heavy modes, $[ \psi \psi' ]$, mixed states of a heavy mode and the volume modulus, $\psi \varphi$, and mixed states of
a heavy mode with the light axion, $ \psi a $. In each case we want to determine the signs of the anomalous dimensions $\gamma_{\psi \psi'}$, $\gamma_{\psi \varphi}$ and $\gamma_{\psi a}$. Following section 4, we restrict the analysis to scalar modes only.

As for the treatment of the light modes,
the underlying parity properties of the $\varphi$ and $a$ fields imply that the relevant Mellin diagram is one involving exchange of the $\varphi$ field -- the $a$ field has odd parity and so a single $a$ field cannot be exchanged. This leaves the $\varphi$ field as the only scalar with low conformal dimension, requiring us 
to determine the $C_{\psi \psi \varphi}$ structure coefficient as the only additional feature
compared to the earlier analysis involving only the light modes.

In a similar fashion to the axion in KKLT, there are two contributions to this. One arises from the kinetic term coupling $f(\varphi) \partial_{\mu} \psi \partial^{\mu} \psi$, and the other from the mass term $m^2(\varphi) \frac{\psi^2}{2}$. The linear coupling can be obtained by deriving the general form of these couplings and then expanding to linear order.

For a general heavy mode $\psi$, we can determine the appearance of the volume modulus in the kinetic and mass terms as follows, by temporarily using the formalism of $\mc{N}=1$ supergravity. As the volume field originates as a K\"ahler modulus, it cannot appear in the superpotential. In an $\mc{N}=1$ supergravity Lagrangian, the kinetic terms and mass fields for the heavy field originate from
\be
K_{\psi \bar{\psi}} \partial_{\mu} \psi \partial^{\mu} \bar{\psi} + e^K \left( K^{\psi \bar{\psi}} D_{\psi} W D_{\bar{\psi}} \bar{W} + \ldots \right) \in \mc{L}.
\ee
The point is that as the K\"ahler moduli $T$ cannot appear in the superpotential, the dependence on the volume has to go via the K\"ahler potential (in particular through the kinetic terms $K_{\psi \bar{\psi}}$). As the overall K\"ahler potential for a IIB compactification is $K = - 2 \ln \mc{V} + \ldots $, the effective Lagrangian for terms quadratic in $\psi$ is
$$
\mc{L}_{\psi\psi} = K_{\psi \bar{\psi}} \partial_{\mu} \psi \partial^{\mu} \bar{\psi} + \frac{1}{\mc{V}^2} K^{\psi \bar{\psi}} \psi \bar{\psi}.
$$
This allows the volume dependence of $K_{\psi \bar{\psi}}$ to be determined, as it is fixed by the requirement that the physical mass
$$
m^2 = \frac{(K^{\psi \bar{\psi}})^2}{\mc{V}^2}
$$
scale correctly as a function of the volume.

The scaling of the physical mass of a heavy mode with the volume (in the large volume limit) is determined on
general principles. As examples, a string mode behaves as
$M_s \propto \frac{M_P}{\sqrt{\mc{V}}}$,
a bulk KK mode behaves as $M_{KK} \propto \frac{M_P}{\mc{V}^{2/3}}$ and a bulk winding mode behaves as
$M_{winding} \propto \frac{M_P}{\mc{V}^{1/3}}$.

For example, for the case of a bulk KK mode, it follows that $K_{\psi \bar{\psi}} \sim \mc{V}^{-1/3}$, and so using
$\varphi = \sqrt{\frac{2}{3}} \ln \mc{V}$, we obtain a kinetic term coupling that scales as $e^{-\sqrt{\frac{1}{6}} \varphi/M_P}$.
There will be additional overall prefactors depending on complex structure (and other) moduli, but as our interest is
in the coupling to the volume mode we can neglect these.

The Lagrangian describing the coupling of the light volume modulus to the heavy bulk KK modes is then
\be
\frac{1}{2} e^{-\sqrt{\frac{1}{6}} (\varphi - \varphi_0)/M_P} \partial_{\mu} \psi \partial^{\mu} \psi
-  m_{\psi}^2 e^{-\frac{5}{\sqrt{6}} (\varphi - \varphi_0)/M_P} \frac{\psi^2}{2}.
\ee

We shift the origin of $\varphi$ by an amount $\varphi_0 = \langle \varphi \rangle$, so that the heavy field $\psi$ is canonically normalised in the vacuum.
Expanding the exponential to first order, the 3-pt couplings are 
\be
\mc{L}_{\varphi \psi \psi} = - \sqrt{\frac{1}{6}} \left( \delta \varphi \right) \frac{\partial_{\mu} \psi \partial^{\mu} \psi}{2} + m_{\psi}^2 \frac{5}{\sqrt{6}} \left( \delta \varphi \right) \frac{\psi^2}{2}.
\ee
The resulting structure function is
\be
\label{ghj}
f_{\varphi \psi \psi} = \frac{\Gamma\left( \frac{2\Delta_{\psi} + \Delta_{\varphi} - 3}{2} \right)}{2 \Gamma(\Delta_{\varphi}) \Gamma(\Delta_{\psi} )^2}
\left( \sqrt{\frac{1}{6}} \left( \Delta_{\varphi} + 2 \Delta_{\psi} - \Delta_{\psi}^2  - 3 \right) + \frac{5}{\sqrt{6}} \Delta_{\psi} \left( \Delta_{\psi} -3 \right) \right).
\ee
As once $\Delta_{\psi} \gg 1$ the quadratic terms in $\Delta_{\psi}$ dominate, this clearly satisfies $f_{\varphi \psi \psi} > 0$ as -- analogously to what occurs in KKLT -- the contribution from the mass term is larger than the contribution from the kinetic term. Naively, this looks reminiscent of the bound discussed in \cite{Andriot:2020lea}, since the sign remains negative as long as the coefficient in the mass term exponential is larger than the $1/\sqrt{6}$ coming from the kinetic term. However, the two numbers are not really independent in this case, and it does not appear possible to draw similar conclusions.

We can ask what condition would lead to opposite sign behaviour, $f_{\varphi \psi \psi} < 0$. This would require the kinetic coupling in Eq. (\ref{ghj}) to be greater than the mass coupling. It is easy to see that the critical volume behaviour here is $K_{\psi \bar{\psi}} = \frac{1}{\mc{V}}$, for which the two terms quadratic in $\Delta_{\psi}$ cancel (the subleading terms are $\mc{O}(1/N)$ suppressed and so are not trustworthy at this order). Physically this would correspond to a state that remains unaltered in mass as $\mc{V} \to \infty$, for example particles arising from D3-branes wrapped on holomorphic 3-cycles.
Positive sign behaviour, $f_{\varphi \psi \psi} < 0$ is then equivalent to having a state that grows in mass in the $\mc{V} \to \infty$ limit, although as previously discussed such `states' would have masses $m_{\psi} > M_P$ and so cannot be regarded as particle states in the effective field theory.

We therefore see that the overall behaviour for the heavy modes can be summarised as
$$
f_{\varphi \psi \psi} > 0 \qquad \equiv \frac{\partial m^2(\psi)}{\partial \varphi} < 0.
$$
$$
f_{\varphi \psi \psi} < 0 \qquad \equiv \frac{\partial m^2(\psi)}{\partial \varphi} > 0.
$$
That is, the sign of the structure function is determined by the behaviour of the massive states: a positive structure function is equivalent to states decreasing in mass with increased volume, and a negative structure function to the opposite.

The anomalous dimension is determined by the (negative) product of the two structure functions,
$$\gamma \propto -f_{\varphi \psi \psi} f_{\varphi \varphi \varphi}$$.
Using the earlier results for LVS, it then follows that a negative anomalous dimension for mixed states involving the heavy modes is equivalent to $\frac{\partial m^2(\psi)}{\partial \varphi} < 0$.
Put another way, a requirement of a negative anomalous dimension is equivalent to requiring heavy states to decrease in mass as the field $\varphi$ moves towards the asymptotic large volume regime of moduli space.
This is interesting as it shows that, within this context, the physics of the distance conjecture can be derived from a simple statement about the sign of anomalous dimensions of mixed double trace operators.

Interestingly, this condition also addresses the puzzle raised above with the refined distance conjecture. Under the
redefinition $\varphi \to - \varphi$, the sign of \emph{both} structure coefficients $C_{\varphi \psi \psi}$ and $C_{\varphi \varphi \varphi}$ (or alternatively $C_{\varphi a a}$) change signs. Although in this Lagrangian, the heavy tower of states all now \emph{increase} in mass for $\Delta \varphi > 0$, as we are headed towards the centre of moduli space, the anomalous dimensions remain negative as they are sensitive to the product of the signs of the structure coefficients. So the formulation in terms of signs of anomalous dimensions appears more fundamental as it captures the correct behaviour in this regime as well.

\subsection{The Presence of the Tower of States}

We therefore see that our hypothetical constraint on the sign of anomalous dimensions for mixed double trace operators captures some of the correct behaviour associated to the swampland distance conjecture. However, the swampland distance conjecture also has a first element - the \emph{existence} of towers of heavy states that becomes light for transPlanckian field displacements. The above constraints on anomalous dimensions constrain the behaviour of the heavy states -- assuming they exist. However, can CFT arguments be used to understand the \emph{existence} of this tower of states?

What does it mean for a tower of heavy states to exist? This can be formulated in terms of the density of operators by conformal dimension, $\rho(\Delta)$. LVS has a small number of single-trace primary states with $\Delta \sim \mc{O}(1)$, as well as other states are built up from these as double-trace (or higher) operators. For conformal dimensions $\Delta \lesssim \mc{V}^{5/6}$ the spectrum of states (on the AdS side) consists of quantum field theory on AdS space, and is built up as the Fock space of states. The analogous behaviour on the CFT side tells us that for $\Delta \lesssim \mc{V}^{5/6}$, $\rho(\Delta)$ grows as for the Fock space of the Generalised Free Field Theory.

However, once $\Delta \gtrsim \mc{V}^{5/6}$, the operators dual to the tower of KK modes enter the spectrum, and for $\Delta \gtrsim \mc{V}$ the operators dual to perturbative string states enter the spectrum. These states are still far below the Planck scale (as $m_s \ll M_P$ in the asymptotic large volume regime), and indeed it is not until $\Delta \gtrsim \mc{V}^{3/2}$ that the CFT will enter the regime in which operators dual to black hole states appear as part of the spectrum. As the number of KK modes grows power-law with energy and the number of string states grows exponentially, this leads to a qualitative change in the functional form of $\rho(\Delta)$ for $\Delta \gtrsim \mc{V}^{5/6}$.

So -- in the context of LVS -- the statement that a tower of heavy states exists is equivalent to the statement that the growth in operator density increases dramatically beyond a certain $\Delta_{crit} \ll \Delta_{BH}$, where $\Delta_{BH}$ is the conformal dimension corresponding to operators dual to black hole states, in a way that no longer can be described by the QFT of single particles on $AdS_4$ space.

Such statements about the spectrum of operator densities have been proven in the context of 2d CFTs  in the context of the idea of string universality \cite{Belin:2014fna} and analogous statements may also hold for higher-dimensional CFTs (see \cite{Alday:2019qrf} for some work in this direction).

\section{Outlook}\label{sc:conc}

Moduli stabilisation is central for reconciling String Theory with experimental observations. Many of the known constructions, however, rely on a series of approximations which may not fully be justified and often involve complicated ingredients that have to be added in `by hand'. In this paper, we have argued that the study of CFT duals to the low energy sector of weakly coupled AdS vacua may help to shed some light on the validity of such constructions, in a way that is complementary to standard arguments. Furthermore, this fits in the recent line of developments concerning the Swampland program, a tentative search of general criteria which low energy theories must satisfy in order to be compatible with Quantum Gravity.

Interestingly, one quantity - the sign of the mixed anomalous dimension $\gamma^{\varphi a}(0,\ell)$, corresponding to double trace operators built out of two non-identical primaries - appears to correlate well with swampland constraints on the effective low-energy Lagrangians for moduli-stabilised string vacua. 
In particular, the requirement that this takes a negative sign appears to reproduce various swampland constraints, both for LVS and the related methods of perturbative stabilisation, as well as for the qualitatively different scenarios of KKLT and racetrack stabilisation. This analysis has also revealed that LVS is rather close to a `critical' point where the sign of the anomalous dimension would change. Furthermore, the same requirement seems to generically relate to the distance conjecture in the large volume limit. In the context of LVS, this is also connected to a possible puzzle within the refined version of the conjecture pointed out in $\ref{ssc:pzl}$.

CFT methods offer promise for rigorous formulations for swampland constraints on semi-realistic string vacua, at least for the first step of AdS vacua. This paper has made some exploratory steps in this direction. However, there remain 
some unclear points which require further investigation:
\begin{itemize}
\item Our focus has been on contributions and anomalous dimensions arising from scalar exchange. Thus gravity appears to be left out, in the 
sense that graviton exchange contributions to the Mellin amplitudes will always dominate at large enough $\ell$ as $\Delta_{g_{\mu nu}} = 3$. One potential approach would be to compute $\gamma(0,\ell)$ at finite values of $\ell$ and verify that the sign is negative for some intermediate finite value $\ell_f \geq 2$, but computations away from the asymptotic limit $\ell \gg 1$ are technically more complicated.

\item For these studied cases of moduli stabilisation, the negativity of the mixed anomalous dimensions arises from exchange of a scalar modulus that corresponds to a volume modulus. Morally, this mode corresponds to a dimensional reduced 10d graviton, and so `ought' to lead to negative anomalous dimensions as gravity is universally attractive. It would be interesting to know whether this extends to more general scenarios, and also under what conditions negativity of the mixed anomalous dimensions occur in arbitrary CFTs.
	
\item Anomalous dimensions of double trace operators can be interpreted as the binding energies of two-particle states on AdS. One might wonder whether this could provide a more transparent interpretation for the negativity of $\gamma^{\varphi a}(0,\ell)$. 

\item It would be interesting if the low-spin behaviour of anomalous dimensions - namely $\gamma(0,0)$ and $\gamma(0,1)$  - could be used to infer something about the quartic vertices, perhaps with some additional assumptions. From a Minkowski perspective, this would be equivalent to asking whether anything can be said on the first two coefficients in the $s$ expansion of a forward scattering amplitude.
\end{itemize}
Aside from these issues, our work can ideally be expanded in two different kinds of directions, corresponding to the trade-off between applicability and rigor that is typical of swampland conjectures. The first possibility consists in testing our speculative criterium for other classes of phenomenologically interesting compactifications, such as type IIA or fibred models. At the other end of the spectrum, it would be useful to investigate precisely to what extent the well-established bounds discussed in \ref{sssc:pos} carry implications for low energy theories of Quantum Gravity on AdS. 
\acknowledgments
The authors are especially grateful to Luis Fernando Alday, Pietro Ferrero and Sirui Ning for fruitful discussions and comments on the manuscript. FR acknowledges support from the Dalitz Graduate Scholarship, jointly established by the Oxford University Department of Physics and Wadham College. JC thanks the KITP, Santa Barbara for hospitality during the completion of the manuscript and the participants of the String Swampland program for various discussions related to this work.

%\begin{thebibliography}{}
   
\providecommand{\href}[2]{#2}\begingroup\raggedright\endgroup

% \end{thebibliography}
% \bibliography{bibabstract}

\end{document}